\documentclass[fleqn,usenatbib]{mnras}

\usepackage{newtxtext,newtxmath}
\usepackage{graphicx}
\usepackage{dcolumn}
\usepackage{bm}
\usepackage{multirow}
\usepackage{color}
\usepackage{changepage}
\usepackage{xcolor}
\usepackage{soul}
\usepackage{physics,amsmath}
\usepackage{ulem}
\usepackage{subcaption}
\usepackage{hyperref}
\usepackage{array}
\usepackage{verbatim}
\usepackage{mathrsfs}
\usepackage{color}
\usepackage{enumitem}
\usepackage{xcolor}
\usepackage{tikz}
\usepackage{orcidlink}

\usepackage[T1]{fontenc}

\DeclareRobustCommand{\VAN}[3]{#2}
\let\VANthebibliography\thebibliography
\def\thebibliography{\DeclareRobustCommand{\VAN}[3]{##3}\VANthebibliography}

\usepackage{graphicx}	
\usepackage{amsmath}	

\title[Tidal dissipation in binary neutron star inspirals]{Tidal dissipation in binary neutron star inspirals from hyperon bulk viscosity: Phase modeling and parameter estimation bias}

\author[Ghosh et al.]{
Suprovo Ghosh,$^{1,2}$\thanks{E-mail: s.ghosh@soton.ac.uk}
Samanwaya Mukherjee,$^{2,3}$
Sukanta Bose$^{4}$
and Debarati Chatterjee$^{2}$
\\
$^{1}$Mathematical Sciences and STAG Research Centre, University of Southampton, Southampton SO17 1BJ, United Kingdom\\
$^{2}$ Inter-University Centre for Astronomy and Astrophysics,
Pune University Campus,
Pune 411007, India\\
$^{3}$International Centre for Theoretical Sciences, Tata Institute of Fundamental Research, Bangalore 560089, India\\
$^{4}$Washington State University, 1245 Webster, Pullman, Washington 99164-2814, USA
}

\date{Accepted XXX. Received YYY; in original form ZZZ}

\pubyear{2025}

\begin{document}
\label{firstpage}
\pagerange{\pageref{firstpage}--\pageref{lastpage}}
\maketitle

\begin{abstract}
During the inspiral of a binary neutron star, viscous processes in the neutron star matter can damp out the tidal energy induced by its companion and convert it to thermal energy. This tidal dissipation/heating process introduces a net phase shift in the gravitational wave signal. In our recent work, we showed based on a Newtonian estimate that tidal dissipation from bulk viscosity originating from the non-leptonic weak interactions involving hyperons could have a detectable phase shift in the gravitational-wave (GW) signal in the next-generation GW detectors. Using simulated signals, we demonstrate that not accounting for this physical effect in waveform models can result in systematic biases in tidal deformability measurements of high-mass neutron star ($\geq 1.8M_{\odot}$) binary observations in next-generation GW detectors. By employing Newtonian orbital dynamics, we model this tidal dissipation induced dephasing as a phenomenological function of the characteristic velocity. We incorporate its effect in gravitational waveforms of equal-mass binary neutron stars. Those waveforms are used to perform a full Bayesian parameter estimation, which confirms that our model can alleviate possible biases in tidal deformability estimation. We also illustrate that the model can accurately measure the additional phase due to tidal dissipation in a $2M_{\odot}$ neutron star in observations with next-generation GW detectors and discuss its significance in extreme matter studies. 
\end{abstract}




\section{Introduction}

Neutron stars (NSs) are unique astrophysical compact objects that can aid our understanding of dense  matter  under  extreme  conditions that are far beyond the reach of terrestrial experiments. Recent multi-messenger and gravitational wave (GW) observations of NSs have facilitated accurate measurements of their masses, radii, and tidal deformabilities. The latter property characterizes the tidal response of NSs during the late stages of a binary inspiral~\citep{Hinderer}, and has been very crucial for studies of dense matter behavior inside neutron stars~\citep{Abbott2017,Annala,De,Most,Dietrich2020,Pang_2021,Issac2021,Huth_2022,Biswas2021,Ghosh_2022,Ghosh2022}. The equation of state (EOS), which represents the behavior of the pressure of the NS matter as a function of its energy density at equilibrium, is essential for calculating these macroscopic properties. Effects of out-of-equilibrium properties of nuclear matter, such as viscosity, on the gravitational wave emission from NSs have also been studied extensively in the context of damping of unstable mode oscillations~\citep{Jones2001,Lindblom2002,Debi_2006,Dalen2004,Jha2010,Haskell2010,Ofengeim,Alford2021,Jha2022} and also recently for the damping of post-merger oscillations~\citep{Alford_2017PRL, Most_2021,Celora_2022,Most_2024,Chabanov_2025,Alford:2019kdw,Alford:2019qtm,Alford:2020lla,Alford:2021lpp,Alford:2022ufz,Alford:2023uih,Alford_2024}.\\

During the inspiral phase of a binary neutron star (BNS) system, tidal interactions of the component stars trigger an exchange of mechanical energy and angular momentum between them at the expense of their orbital energy. These tidal interactions can drive the system out of equilibrium depending on the relevant nuclear reactions timescales. These out-of-equilibrium viscous processes inside the star damp out the tidal energy and convert this energy to heat which we refer to as ``tidal dissipation" or equivalently ``tidal heating". The tidal dissipation also induces a ``tidal lag" angle between the direction of the bulge and the orbital separation. At low temperatures (T $\leq 10^9$ K) relevant to the inspiral phase of a BNS coalescence, the dominant source of this dissipation is the shear viscosity, 
arising from the momentum transport due to $ee$ and $nn$ scattering~\citep{Sawyer1989,Lai_1994}. Earlier,~\cite{Lai_1994} found that shear viscous damping of mode oscillations of NSs during the inspiral could only heat the stellar core to T $\sim 10^8$K, and the timescale of the viscous dissipation being much longer than the inspiral timescale, the imprints of these effects on the dynamics of a BNS merger are negligible in gravitational-wave studies~\citep{Bildsten1992,Lai_1994}. For this reason, although extensive studies exist in the literature on developing a relativistic theory of static and dynamical tides for binary neutron star inspiral~\citep{Hinderer,Poisson:2020vap,Pitre:2023xsr,TEOB,NR,Bernuzzi_TEOB}, the tidal dissipation of neutron stars has been studied mostly in Newtonian gravity~\citep{Lai_1994,Ripley_2023,Ghosh_2024}. Recently, in ~\cite{HegadeKR:2024slr}, signatures of the tidal lag in gravitational waves were re-analyzed including relative 1PN effects. In ~\cite{HegadeKR:2024agt}, an analysis of relativistic dynamical tidal response for neutron stars was performed by obtaining a resummed all-orders-in-frequency tidal response to linear order in viscosity to study both the conservative and dissipative effects. A recent study by ~\cite{Saketh_2024} also presented  a theory of tidal heating in neutron stars in a fully relativistic formalism as a gravitational Raman scattering problem.  In these recent works based on tidal lag~\citep{Ripley_2024,HegadeKR:2024agt,Saketh_2024}, it was assumed that the tidal lag (termed as ``dissipation number") remains constant throughout the binary inspiral timescale. However, this assumption fails in realistic scenarios since the viscous coefficients are dependent on the local temperature profile inside the NS, which heats up during the inspiral due to tidal dissipation, making tidal lag a dynamical parameter.
\\

These earlier studies in Refs.~\citep{Lai_1994,Arras2019} considered the dominant source of viscosity to be originating from ordinary nucleonic matter inside neutron stars. But at the core of the neutron stars, strangeness containing exotic particles, such  as hyperons, kaons or even deconfined quark matter, can become stable components due to weak equilibrium~\citep{GlendenningBook,Lattimer2004,Blaschke2018}. Although bulk viscosity originating from direct and modified Urca reactions are dominant at high temperatures (T $\geq 10^9$ K), hyperonic bulk viscosity originating from non-leptonic processes can be several orders higher ($\approx 10^{8} - 10^{10}$ times) than the shear viscosity from $ee$ scattering in the temperature range of $10^{6}-10^8$K~\citep{Lindblom2002,Debi_2006,Alford2021}. Recently, we have shown in Ref.~\citep{Ghosh_2024} that tidal heating due to the dissipation from hyperon bulk viscosity can heat up the star up to $0.1-1$ MeV, a range much higher than the earlier estimates. Based on a leading order phase estimation, we found the phase difference induced in the GW signal due to this dissipation to be of the order $10^{-3} - 0.5$ rad, depending on the component neutron star masses.\\

With the increased sensitivity of the future third-generation GW detectors such as Einstein telescope (ET)~\citep{ET1,ET2} in Europe and the Cosmic Explorer (CE)~\citep{CE} in USA, we expect to see many BNS merger events with some of the events reaching a signal-to-noise ratio (SNR) of 500 or more~\citep{Iacovelli:2023nbv,Abac:2025saz}. In order to assess the detectability of the small shifts due to tidal dissipation in these next generation GW detectors, one needs to accurately model the phase difference introduced into the waveform. Since there is no existing state-of-the-art BNS waveform model incorporating tidal dissipation effects, we consider the Newtonian tidal dissipation estimates based on our earlier study~\citep{Ghosh_2024}, in which we obtained the phase difference using Newtonian orbital energy. The current work extends that study by calculating the additional phase via the stationary phase approximation (SPA)~\citep{Tichy:1999pv}, while also including post-Newtonian corrections. We then incorporate this additional phase correction into existing BNS waveforms and perform Bayesian parameter estimation of simulated events in third-generation GW detectors, such as the Einstein Telescope (ET)~\citep{ET1,ET2} and Cosmic Explorer (CE)~\citep{CE} to investigate both the detectability of this additional phase and possible biases in the recovered tidal deformability parameter in current waveform models. To capture the additional dependence of the dissipation parameter on the frequency, we model the frequency-domain GW phase as a function of orbital velocity that best fits the GW phase correction.  Then, we also perform Bayesian parameter estimation studies to evaluate its effect on the measurement of NS tidal deformability. 
\\

The article is organized as follows: in Sec.~\ref{sec:HyperonBV}, we calculate the energy dissipated due to the hyperon bulk viscous dissipation of the dominant $f$-mode. In Sec.~\ref{sec:Phase}, we estimate the phase due to this energy dissipated and estimate the bias in the recovery of tidal deformability from binary NS mergers using current and future generational detectors. In Sec.~\ref{sec:model}, we model the frequency domain phase of the waveforms using polynomial functions of orbital velocity. We also perform full Bayesian parameter estimation studies to show how accurately we can determine this additional phase from simulated events in future-generation detectors. Finally, in Sec.~\ref{sec:discussion} we discuss the main implications of this work and also future directions. We use the geometric units, assuming $G=c=1$, unless stated explicitly otherwise.

\section{Dissipated tidal energy in the mode-sum approach}\label{sec:HyperonBV}
In this section we recapitulate the theory of Newtonian tidal heating from linear perturbations of a background solution for a star in equilibrium, following Ref.~\citep{Ghosh_2024,Lai_1994}. Under the adiabatic approximation, the effect of the tidal potential due to the companion star is measured in terms of the Lagrangian fluid displacement vector $\bm{\xi}(r,t)$ from its equilibrium position. This displacement can be analysed in terms of the normal modes of the neutron star, 
\begin{equation}\label{eq:normal}
    \bm{\xi} (\bm{r},t) = \sum_{\alpha}\bm{\xi_{\alpha}(r)} a_{\alpha}(t)\,,
\end{equation}
where $\alpha \equiv \{n,l,m\}$ denotes the normal mode index, $\bm{\xi_{\alpha}(r)}$ is the eigenfunction and $a_{\alpha}(t)$ is the time-dependent amplitude of the particular eigenmode due to the tidal field of the companion. During the inspiral of the binary neutron star system, tidal interactions may induce resonant or non-resonant excitation of these oscillation modes inside the star depending on their frequencies ($\omega_{\alpha}$) compared to the orbital frequency ($\Omega$)~\citep{Andersson_2018}. Since the fluid is viscous, a fraction of this dynamical tidal energy is dissipated as thermal energy, and increase the temperature of the system. This energy dissipation depends on the timescale of viscous dissipation compared to the orbital timescale. Assuming the viscous dissipation timescale is much larger than the inspiral timescale, the time-dependent amplitude of a particular mode is governed by the equation~\citep{Lai_1994}
\begin{equation}\label{eq:mod_evol}
    \Ddot{a}_{\alpha} + \gamma_{\alpha}\Dot{a}_{\alpha}+ \omega^2_{\alpha}a^2_{\alpha} = -\frac{M'W_{lm}Q_{nl}}{D^{l+1}}e^{i\Phi(t)}\,,
\end{equation}
where $M'$ is the companion mass, $Q_{nl}$ is the tidal coupling for the mode, $\gamma_{\alpha}$ is the viscous damping rate, $D$ and $\Phi$ are the separation and phase of the decaying orbit respectively, and $W_{lm}$ are numerical coefficients defined as
\begin{equation}
\begin{split}
        W_{lm} &= \left(-1\right)^{-(l+m)/2}\left[\frac{4\pi}{2l+1}(l+m)!(l-m)!\right] \\& \times \left[2^l\left(\frac{l-m}{2}\right)!\left(\frac{l+m}{2}\right)!\right]^{-1}\,.
\end{split}
\end{equation}
The viscous damping rate (or equivalently the inverse of dissipation timescale) for these normal modes is given  by
\begin{equation}\label{eq:damptime2}
    \gamma_{\alpha} = \Dot{E}_{{\rm visc},\alpha}/2E_{\alpha}\,,
\end{equation}
where $E_{\alpha}$ is the energy of the mode and $\Dot{E}_{{\rm visc},\alpha}$ is the energy dissipation rate. For a viscous fluid, the rate of dissipated energy is given in terms of the viscous stress tensor $\sigma_{ij}$~\citep{Lai_1994}
\begin{equation}\label{eq:viscenergy}
    \Dot{E}_{\rm visc} = \int d^3x \sigma_{ij}\bm{V}_{i,j}\,,
\end{equation}
where $\bm{V}$ denotes the perturbation velocity vector. The viscous stress tensor $\sigma_{ij}$ can be written as~\citep{Landau1987Fluid}
\begin{equation}\label{eq:visctensor}
    \sigma_{ij} = \eta_{SV}\left(\bm{V}_{i,j} + \bm{V}_{i,j} - \frac{2}{3}\delta_{ij}\nabla .\bm{V}\right) + \zeta\delta_{ij}\nabla . \bm{V}\,,
\end{equation}
where $\eta_{SV}$\footnote{subscript added to avoid any confusion with the symmetric mass ratio of a binary system used later.} and $\zeta$ are the shear and bulk viscosity coefficients respectively. For any particular mode, the eigenfunction can be written as a sum of the radial and tangential components: 
\begin{equation}\label{eq:eigenvector}
    \bm{\xi_{\alpha}}(r) = \left[\xi^r_{nl}(r)\bm{e_r} + r\xi^{\perp}_{nl}(r)\bm{\nabla}\right]Y_{lm}(\theta,\phi)\,,
\end{equation}
where $\bm{e_r}$ is the radial vector and $Y_{lm}(\theta,\phi)$ are the spherical harmonic functions. Using the expression of the displacement vector in Eq.~\eqref{eq:normal} and $ \bm{V} = \frac{d}{dt}\bm{\xi} (r,t)$ in Eq.~\eqref{eq:viscenergy}, we get the viscous dissipated energy as 
\begin{equation}\label{eq:viscenergy_final}
    \Dot{E}_{\rm visc} \approx \sum_{\alpha}2\gamma_{\alpha}\Dot{a}_{\alpha}(t)^2
\end{equation}
in this mode-sum approach during the inspiral. To leading order, when the viscous dissipation rate is smaller than twice the mode frequency($\gamma_{\alpha} < 2\omega_{\alpha}$), the mode  amplitude $a_{\alpha}$ can be obtained by solving Eq.~\eqref{eq:mod_evol} and then plugging it into Eq.~\eqref{eq:viscenergy_final} to get the total dissipated energy.\\

\begin{figure}
    \centering
    \includegraphics[width=0.48\textwidth]{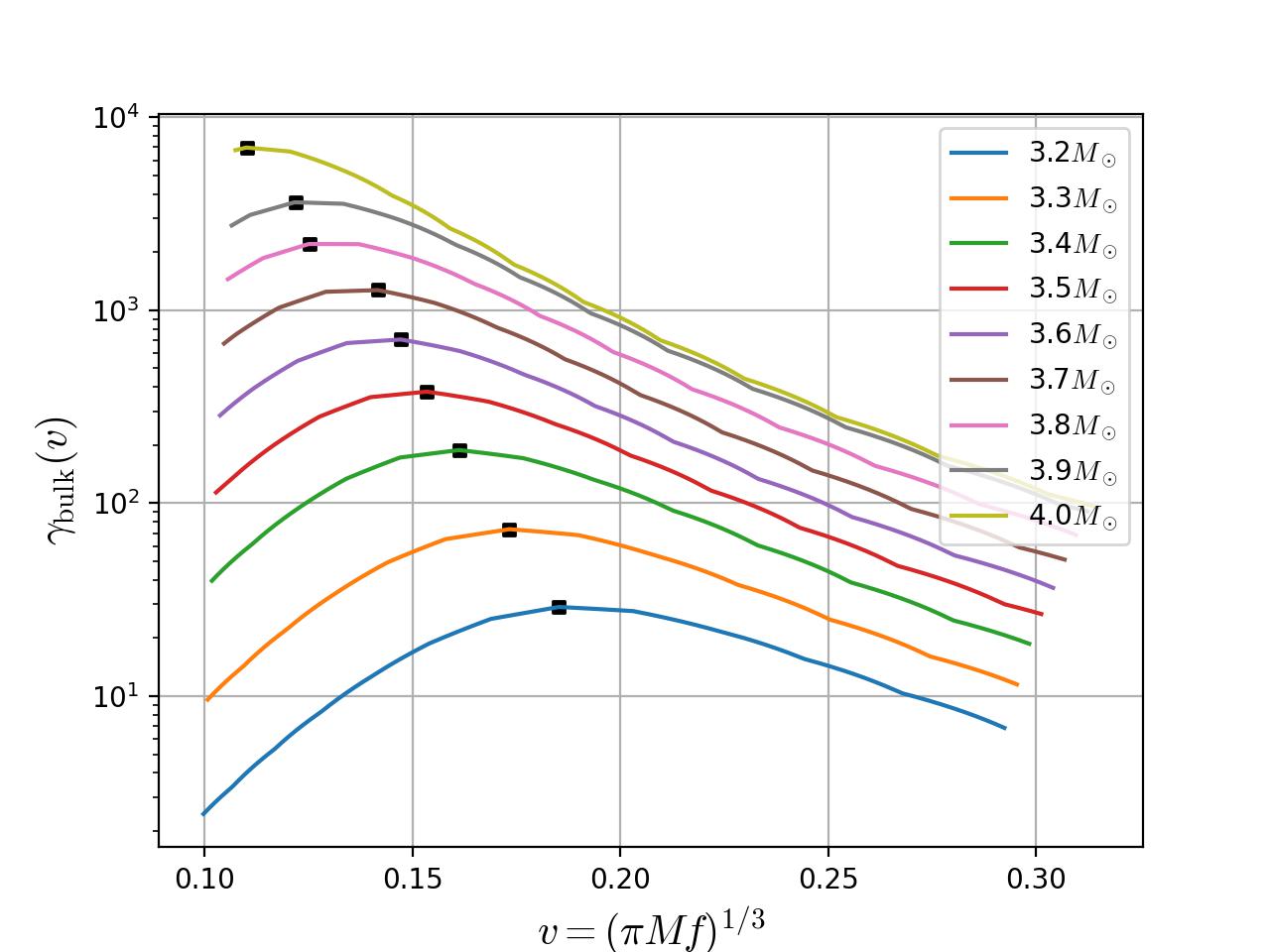}
    \captionsetup{justification=centerlast}
    \caption{\small Bulk viscous dissipation rate of hyperonic neutron stars as a function of their characteristic velocity in gravitational-wave binaries. Equal-mass BNS systems are considered here, with the legend reporting the total mass for each of the binaries. The black squares represent the individual maxima. }
    \label{fig:gammavsv}
\end{figure}

In this work, we are considering the energy dissipated due to the viscous damping of the dominant \textit{f}-mode. Although there are other modes such as low frequency $g$-modes that can contribute to tidal heating, the coupling of these modes to the GW emission is very small~\citep{Andersson_2018}, rendering their effects largely subdominant. Given a background equilibrium EOS, we determine the $f$-mode frequency and eigenfunctions via a relativistic Cowling approximation~\citep{Pradhan_2021} and the normalised mode eigenfunctions are also used to calculate the tidal coupling defined as 
\begin{equation}
   Q_{nl} = \int_0^R \rho lr^{l+1} [\xi^r_{nl}(r)\bm{e_r} + r\xi^{\perp}_{nl}]dr\,,
\end{equation}
$R$ being the radius of the star and $\rho$ its energy-density. Also, we only consider the dissipation that comes from the bulk viscosity originating from the weak nonleptonic processes involving $\Lambda$ hyperons, since this has been shown to have a detectable effect during the binary inspiral~\citep{Ghosh_2024}. If we consider only the bulk viscosity contribution to the energy dissipation given in Eq.~\eqref{eq:viscenergy} for the $f-$mode, we can express the viscous dissipation rate as~\citep{Lai_1994}
\begin{align} \label{eq:gammabulk}
    \gamma_{\rm bulk} =  \frac{1}{2} \frac{(l+|m|)!}{(l-|m|)!}\int_0^R r^2dr\zeta&\left(\frac{\partial \xi^r}{\partial r} + \frac{2}{r}\xi^r\right. \nonumber\\& \left.- l(l+1)\frac{\xi^{\perp}}{r}\right)^2\,, 
\end{align}
where $\xi^r$ and $\xi^{\perp}$ are the radial and perpendicular component of the $f$-mode eigenfunction, and the corresponding velocity field can be written as  $\bm{v} = -i\omega \bm{\xi}$. Given an equilibrium EOS and the bulk viscosity from weak interactions, we have shown in Ref.~\citep{Ghosh_2024} how the temperature changes as a function of separation between the binary masses which is related to the inspiral frequency. Having obtained the temperature as a function of the inspiral frequency, one can relate the latter with bulk viscosity and integrate Eq.~\eqref{eq:gammabulk} inside the star to get $\gamma_{\rm bulk}$ as a function of the orbital velocity. In Fig~\ref{fig:gammavsv}, we show this for different binary systems of equal mass for the parametrization of the FSU2 EOS as considered in~\cite{Ghosh_2024}. Later, we also consider other EOS parametrizations, spanning the allowed uncertainty range in the EOS while ensuring compatibility with astrophysical data. The resonance-like behavior of $\gamma_{\rm bulk}$ with the orbital velocity comes from the resonance of hyperon bulk viscosity - matching of relevant reaction rates to the perturbation timescale at a finite temperature (refer to Fig. 3 in ~\cite{Ghosh_2024}). The increase in the $\gamma_{\rm bulk}$ values with higher masses is due to the increase in hyperon fractions inside the heavier neutron stars. \\

Considering the amplitude of the $l = m =2$ $f$-mode obtained by integrating Eq.~\eqref{eq:mod_evol}, the viscous energy dissipation rate in an equal-mass binary system was estimated to be~\citep{Ghosh_2024}
\begin{equation}\label{eq:viscener}
    \Dot{E}_{\rm visc} = \frac{24\pi}{5}\frac{M^2}{R}\omega_0^{-4}Q_0^2\left(\frac{R}{D}\right)^9\gamma_{\rm bulk}\,,
\end{equation}
where $M$ is the total mass, $Q_0$ is the tidal coupling strength of the $f$-mode and $\omega_0$ is the normalised frequency of the $f$-mode.

\section{Estimation of phase \& effect on waveform}\label{sec:Phase}

\subsection{Numerical dephasing calculation}\label{sec:phase_calc}
The dissipative loss of energy and angular momentum in a hyperonic NS due to its high bulk viscosity drains energy from its orbit during the inspiral of a BNS coalescence, resulting in a faster inspiral rate. The latter fact introduces changes in the phase evolution of the gravitational waveforms of these binaries as predicted by general relativity (GR). To construct gravitational waveforms for a compact binary coalescence (CBC) under GR, the Einstein field equations (EFEs) can be analytically solved, in a perturbative manner, under the post-Newtonian (PN) framework. PN formalism works well when the system under consideration can be approximated to be weakly gravitating and its components slowly moving in the center-of-mass frame. For a BNS system, these conditions translate to the requirement that the NSs orbit each other with velocities much lower than the speed of light in vacuum, and they are sufficiently far apart so that the system is not too compact. In this formalism, the evolution of the orbital phase $\phi(t)$ of a compact binary system is computed as a perturbative expansion in a small parameter, typically taken to be the characteristic velocity $v = (\pi M f)^{1/3}$, $M$ being the total mass of the binary. This analytical procedure requires $v\ll 1$, which makes it useful in the early inspiral phase of a CBC.

The loss of binding energy $E(v)$ of the two-body system with time equals the GW flux emitted to future null infinity ($\mathcal{F}^{\infty}(v)$) plus the energy dissipated due to the internal viscous forces of NS ($\dot{E}_{\rm visc}(v)$). So the energy balance condition becomes
\begin{equation}\label{eq:energybalance}
    -\dv{E(v)}{t}=\mathcal{F}^{\infty}(v)+\dot{E}_{\rm visc}(v).
\end{equation}

Evolution of the orbital phase $\phi$ and the characteristic velocity $v$, obtained from this equation, read
\begin{equation}
    \dv{\phi}{t} = \frac{v^3}{M}\,,\quad\quad\quad \dv{v}{t}=-\frac{\mathcal{F}(v)}{E'(v)},
\end{equation}
where $\mathcal{F}(v)=\mathcal{F}^{\infty}(v)+\dot{E}_{\rm visc}(v)$. These equations yield a solution for the phase $\Phi(f)$ of the frequency-domain waveform $\Tilde{h}(f) =  \Tilde{A}(f) e^{-i\Phi(f)}$~\citep{Tichy:1999pv}:
\begin{equation}
\label{psi}
    \Phi(f)=2(t_c/M)v^3 - 2\phi_c - \pi/4 - \frac{2}{M}\int (v^3-\Bar{v}^3)\frac{E'(\Bar{v})}{\mathcal{F}(\Bar{v})}\dd\Bar{v},
\end{equation}
where $E'(v)=\dv*{E(v)}{v}$. 
The separation $D$ between the NSs in Eq.~\eqref{eq:viscener} and the orbital frequency $\Omega$ are related by
\begin{equation}
    \Omega^2=\frac{M}{D^3}\,,
\end{equation}
where $\Omega=\pi f$, $f$ being the GW frequency corresponding to the (2,2) mode. So we get
\begin{equation}
    (\pi M f)^2=\frac{M^3}{D^3}\,,
\end{equation}
implying
\begin{equation}
    D(v)=\frac{M}{v^2}\,.
\end{equation}
Using this in Eq.~\eqref{eq:viscener} we get (in geometric units)
\begin{equation}
    \dot{E}_{\rm visc}\propto (M^2/R)\times(Q_0^2/\omega_0^4)\times(R/M)^9\times(\gamma_{\rm bulk})\times v^{18}\,.
\end{equation}
Defining the compactness to be $C=M/R$, this can be written as
\begin{equation}
    \dot{E}_{\rm visc}\propto C^{-7}\times(Q_0^2/\omega_0^4)\times\gamma_{\rm bulk}(v)\times v^{18}\times R\,.
\end{equation}

\begin{figure}
    \centering
    \includegraphics[width=0.5\textwidth]{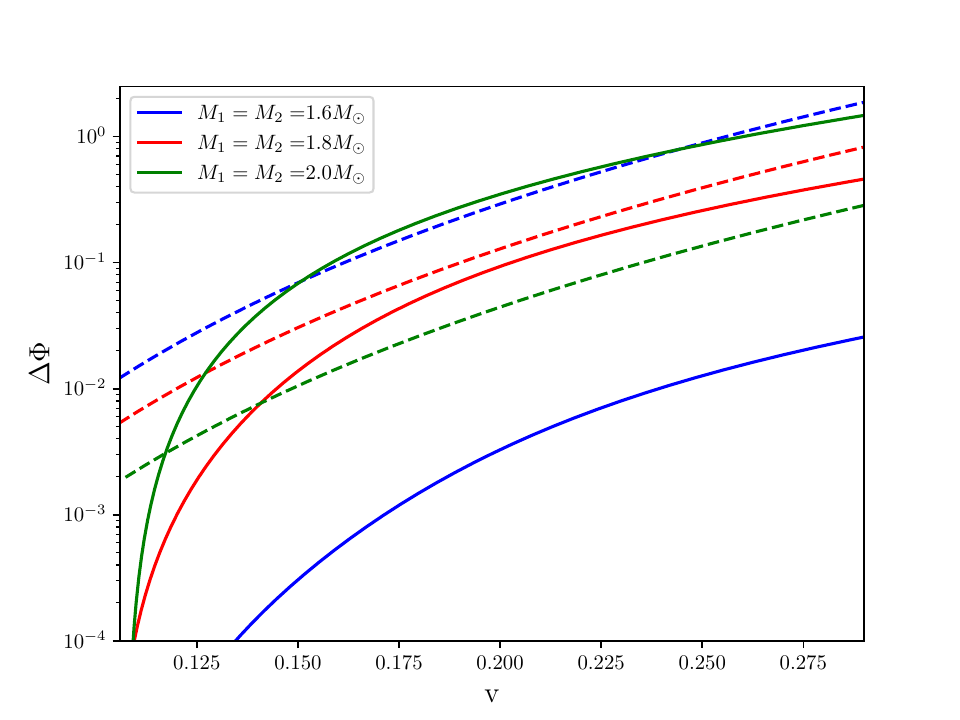}
    \captionsetup{justification=centerlast}
    \caption{Estimated phase due to tidal deformability (dotted lines) and tidal heating (solid lines) as a function of $v = (\pi Mf)^{1/3}$ for equal mass binary of $1.6, 1.8$ and $2.0 M_{\odot}$ individual masses.}
    \label{fig:fig_3}
\end{figure}

Up to the leading order (LO) and next-to leading order (NLO), the post-Newtonian expansions for the functions $E(v)$ (orbital  energy) and $\mathcal{F}^{\infty}(v)$ (energy flux to the infinity)  have the general form~\citep{Isoyama:2017tbp}
\begin{equation}\label{EV}
    E(v) = -\frac{1}{2}\eta M v^2\left[1 - \frac{(9+\eta)}{12}v^2\right]\,,
\end{equation}
and 
\begin{equation}\label{FV}
    \mathcal{F}^{\infty}(v) = \frac{32}{5}v^{10}\eta^2\left[1 - v^2\left(\frac{1247}{336}+\frac{35\eta}{12}\right) + 4\pi v^3\right]\,.
\end{equation}
where $\eta$ is the symmetric mass ratio of the binary system defined as $\eta = \frac{M_1M_2}{(M_1 + M_2)^2}$. We plug the expressions for $E(v)$, $\mathcal{F}^{\infty}(v)$ and $\Dot{E}_{\rm visc}$ from Eqs.~\eqref{EV},~\eqref{FV} and~\eqref{eq:viscener} respectively in Eq.~\eqref{psi}, and integrate to get the phase of the gravitational waveform. To see the relative phase difference due to the tidal dissipation alone, we simply subtract the numerically integrated phases obtained by integrating Eq.~\eqref{psi} with and without considering the tidal dissipation. In Fig.~\ref{fig:fig_3}, we plot this numerically obtained phase difference due to the tidal dissipation for three equal-mass binary systems considering the FSU2 EOS parametrizations from ~\cite{Ghosh_2024}.

\begin{figure*} 

\begin{subfigure}{0.32\textwidth}
\includegraphics[width=\linewidth]{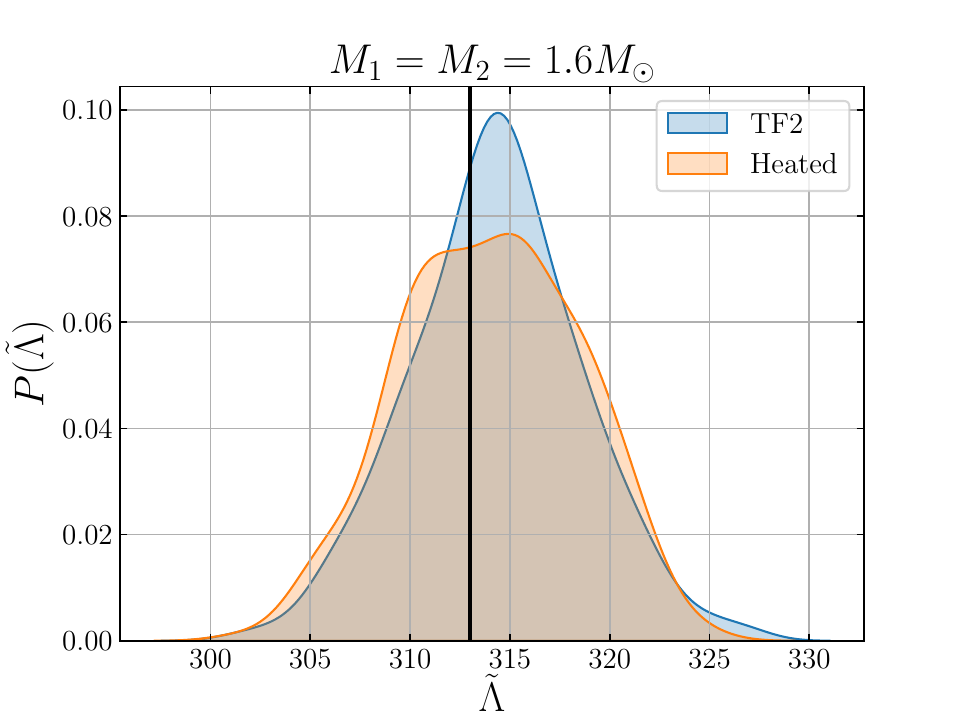}
\end{subfigure}\hspace*{\fill}
\begin{subfigure}{0.32\textwidth}
\includegraphics[width=\linewidth]{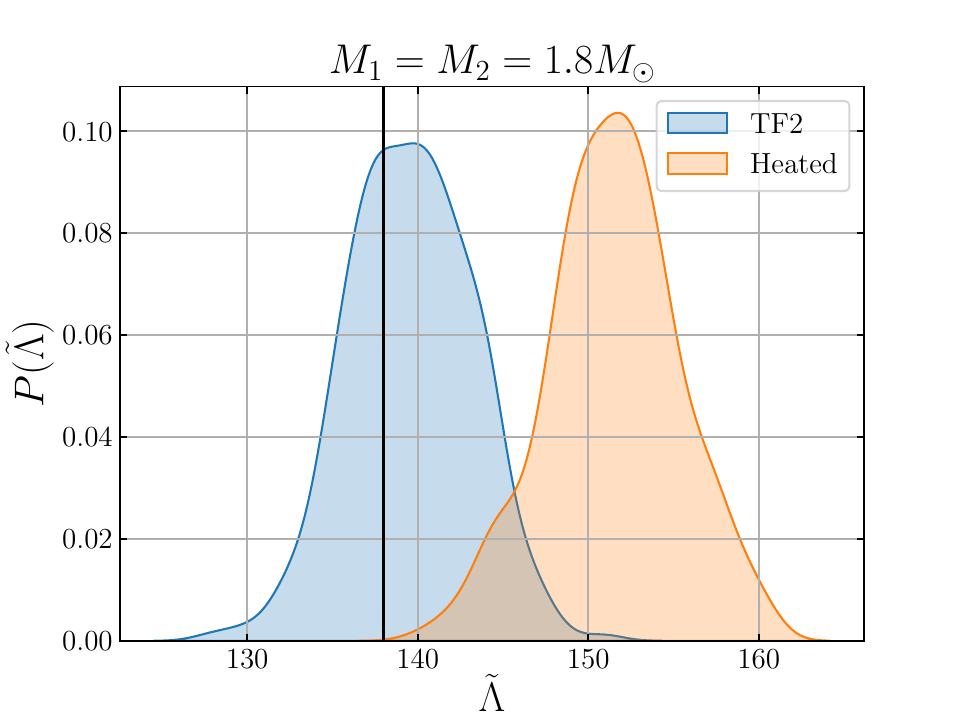}
\end{subfigure}\hspace*{\fill}
\begin{subfigure}{0.32\textwidth}
\includegraphics[width=\linewidth]{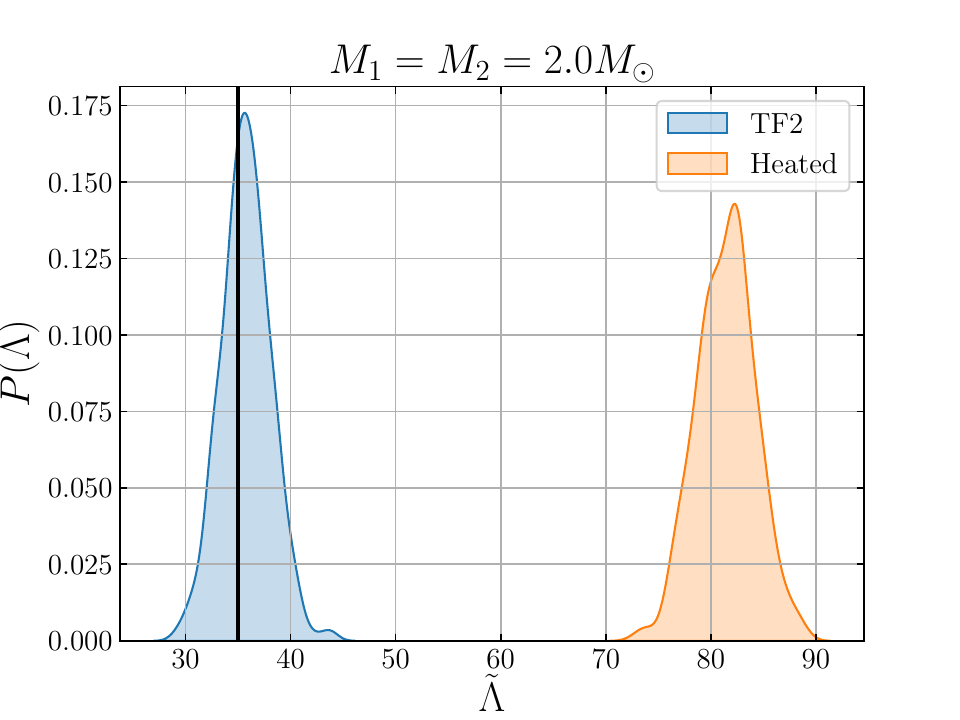}
\end{subfigure}

\medskip
\begin{subfigure}{0.32\textwidth}
\includegraphics[width=\linewidth]{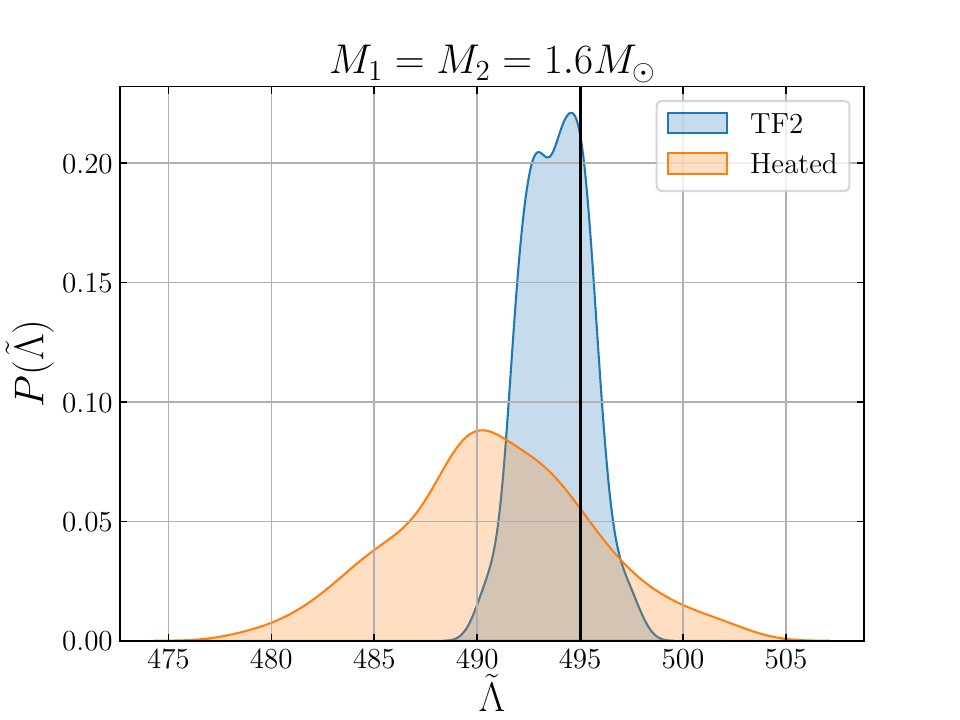}
\end{subfigure}\hspace*{\fill}
\begin{subfigure}{0.32\textwidth}
\includegraphics[width=\linewidth]{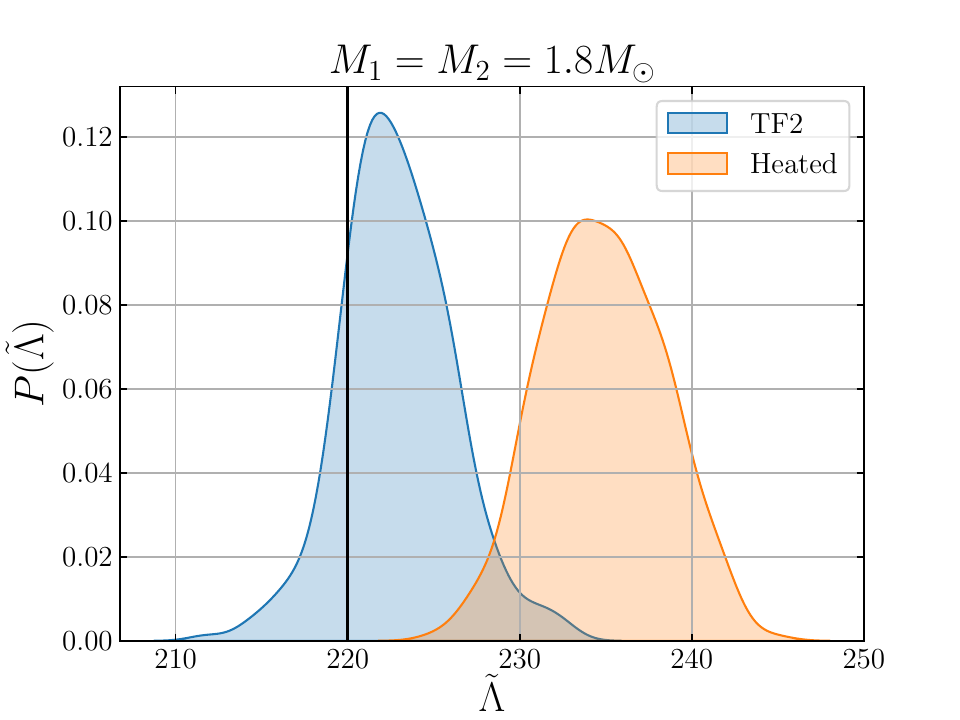}
\end{subfigure}\hspace*{\fill}
\begin{subfigure}{0.32\textwidth}
\includegraphics[width=\linewidth]{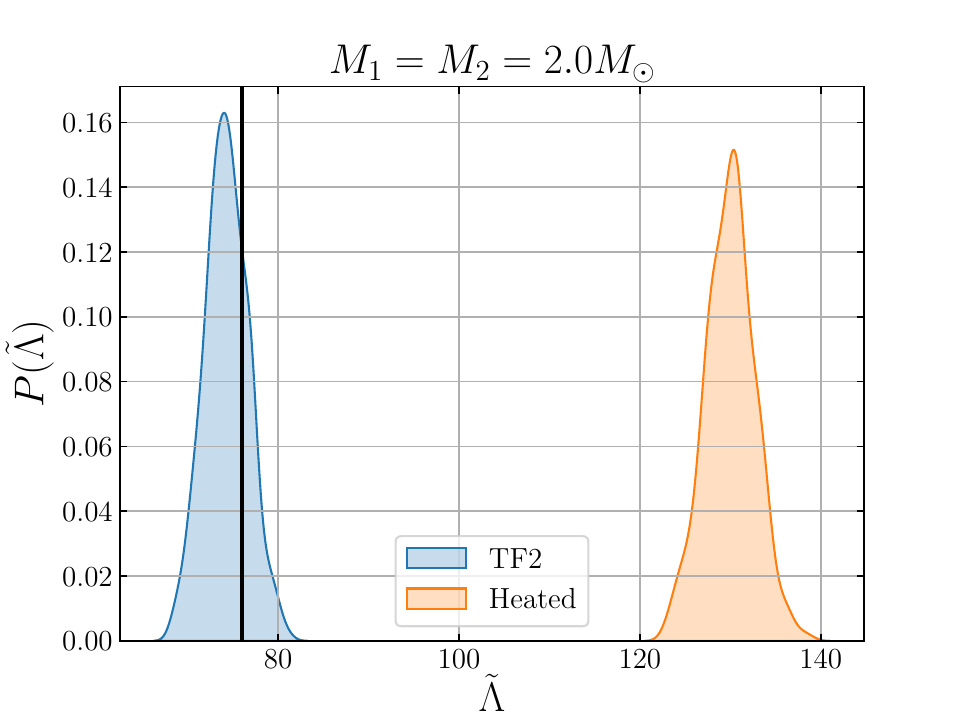}
\end{subfigure}

\medskip
\begin{subfigure}{0.32\textwidth}
\includegraphics[width=\linewidth]{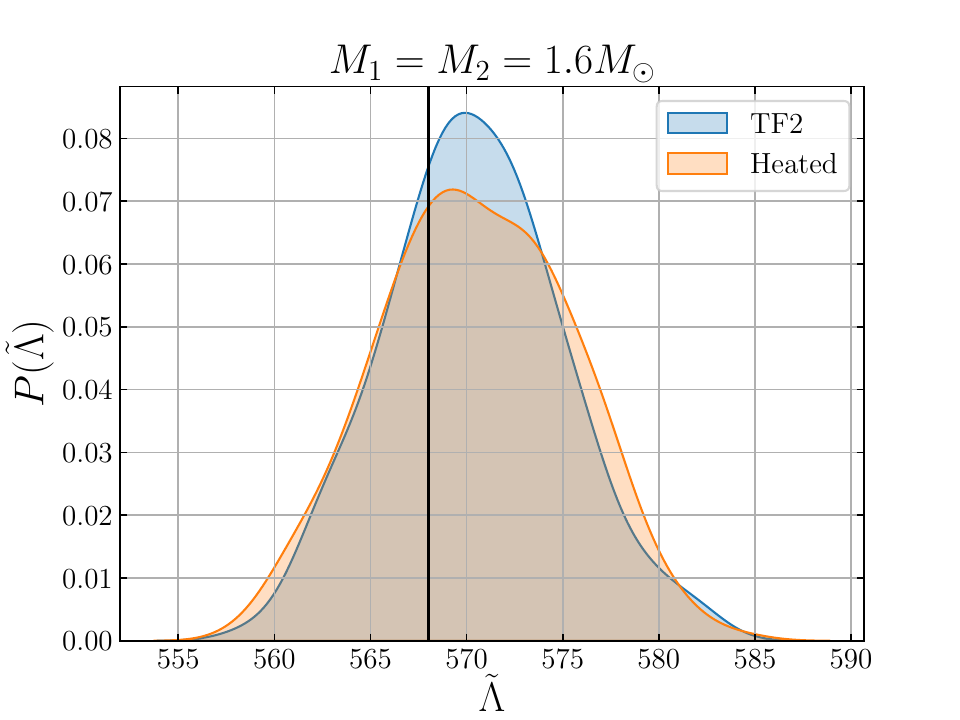}
\end{subfigure}\hspace*{\fill}
\begin{subfigure}{0.32\textwidth}
\includegraphics[width=\linewidth]{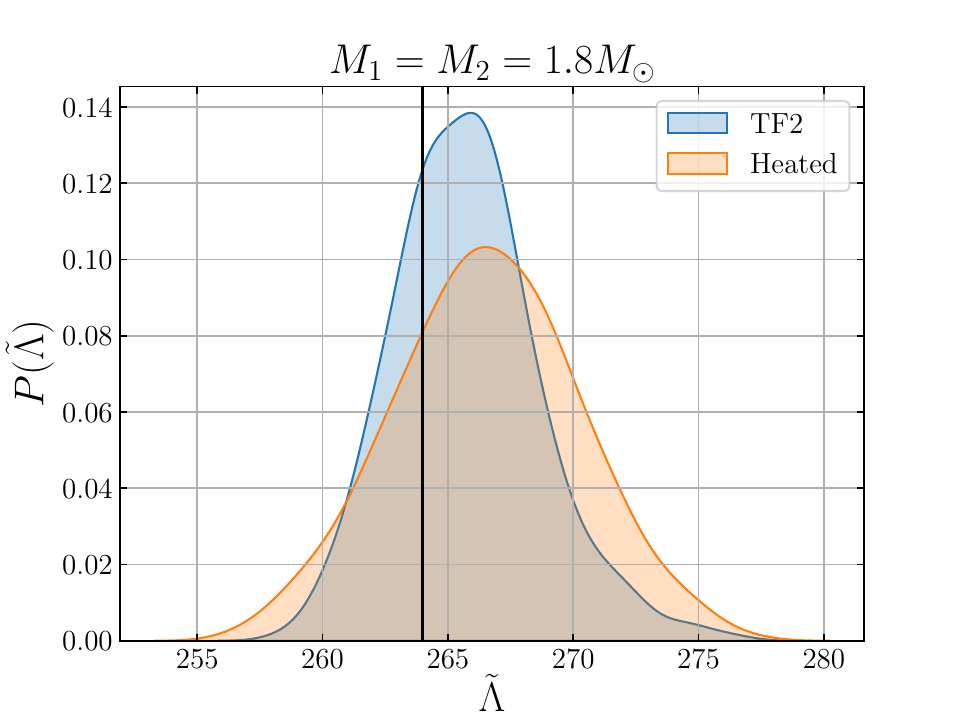}
\end{subfigure}\hspace*{\fill}
\begin{subfigure}{0.32\textwidth}
\includegraphics[width=\linewidth]{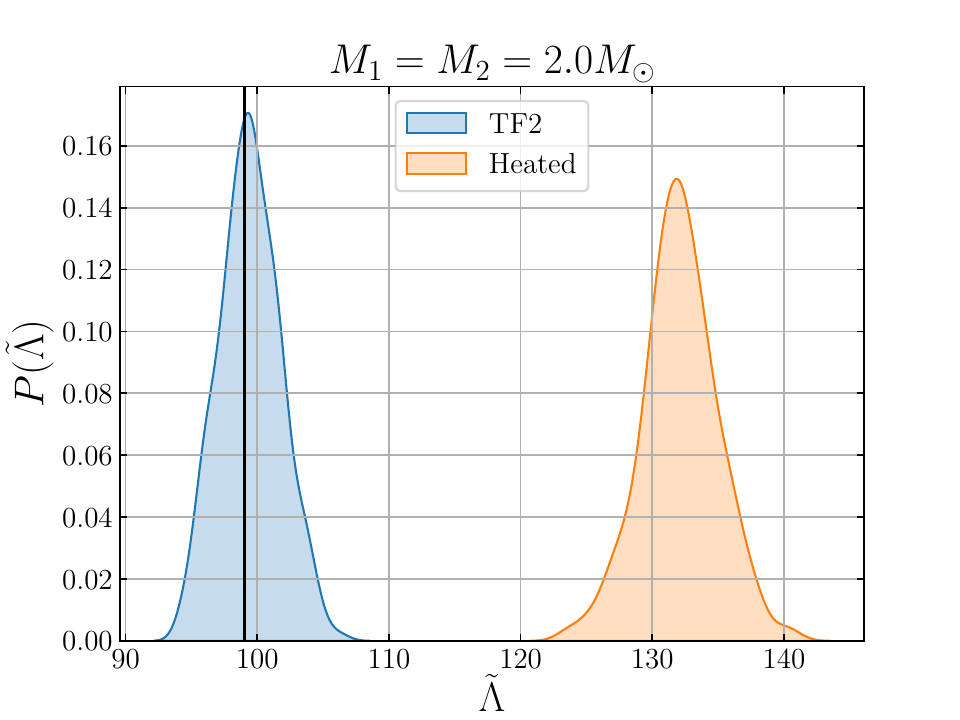}
\end{subfigure}

\caption{Injection and recovery of effective tidal deformability from BNS event with ET-D sensitivity of equal masses for three different EOSs: a) HZTCS(upper panel) b)FSU2(middle panel) and c)Nl3(lower panel). Blue and orange posteriors show injection without(TF2) and with effects of tidal heating(Heated) respectively. Recovery is always done with the TF2 model. Black line shows the injected value. } 

\label{fig:fig_5}
\end{figure*}

\subsection{Injection-Recovery Study for bias in Tidal deformability}
\label{sec:Bias}
To see how this extra phase shift due to the tidal dissipation which is not accounted for by any current BNS waveform model can impact the inference of the neutron star properties from the binary mergers, we compare this numerically obtained dephasing to what we can expect from the leading order tidal deformability. The Newtonian or leading order dephasing due to the tidal deformability enters the GW waveform at 5PN order compared to the point-particle phase and given by~\citep{Hinderer}
\begin{equation}\label{eq:phaseTD}
    \delta\Phi_{TD} =  - \frac{117\tilde{\lambda}v^5}{8\eta M^5}
\end{equation}
where $\tilde{\lambda}$ is defined as 
\begin{equation}
    \tilde{\lambda} = \frac{1}{26}\left[\frac{M_1+12M_2}{M_1}\lambda_1 + \frac{M_2+12M_1}{M_2}\lambda_2\right].
\end{equation}
Given an EOS, the tidal deformability parameter ($\lambda$) can be calculated  by  solving  a  set  of  differential  equations coupled  with  the  TOV  equations~\citep{Hinderer} and is related to the $l = 2$ love number ($k_2$) as
\begin{equation}
    \lambda = \frac{2}{3} k_2R^5.
\label{eq:love}
\end{equation}
In Fig~\ref{fig:fig_3}, for three equal mass binary systems we have compared the numerically obtained phase difference as described in Sec.~\ref{sec:phase_calc} due to tidal dissipation with same from tidal deformability from eqn.~\eqref{eq:phaseTD}~\citep{Hinderer}. For neutron stars of mass $\geq 1M_{\odot}$, the dimensionless tidal deformability decreases with increasing mass~\citep{Hinderer}. As a result, the magnitude of dephasing also decreases with increasing component masses of the binary system. On the other hand, increasing component masses entail a higher hyperon content inside the star(Refer to Table I in ~\cite{Ghosh_2024}) leading to more tidal dissipation~\citep{Ghosh_2024}. We see that for $1.6M_{\odot}$, the dephasing due to tidal deformability is higher than due to tidal heating. For $1.8M_{\odot}$, the dephasing for both are almost same and for very high mass of $2M_{\odot}$, the dephasing due to tidal heating dominates. So, we expect to see large systematic biases in estimation of tidal deformability from these high mass systems if the effects of tidal dissipation are not accounted for properly.

\begin{figure*} 
\begin{subfigure}{0.32\textwidth}
\includegraphics[width=\linewidth]{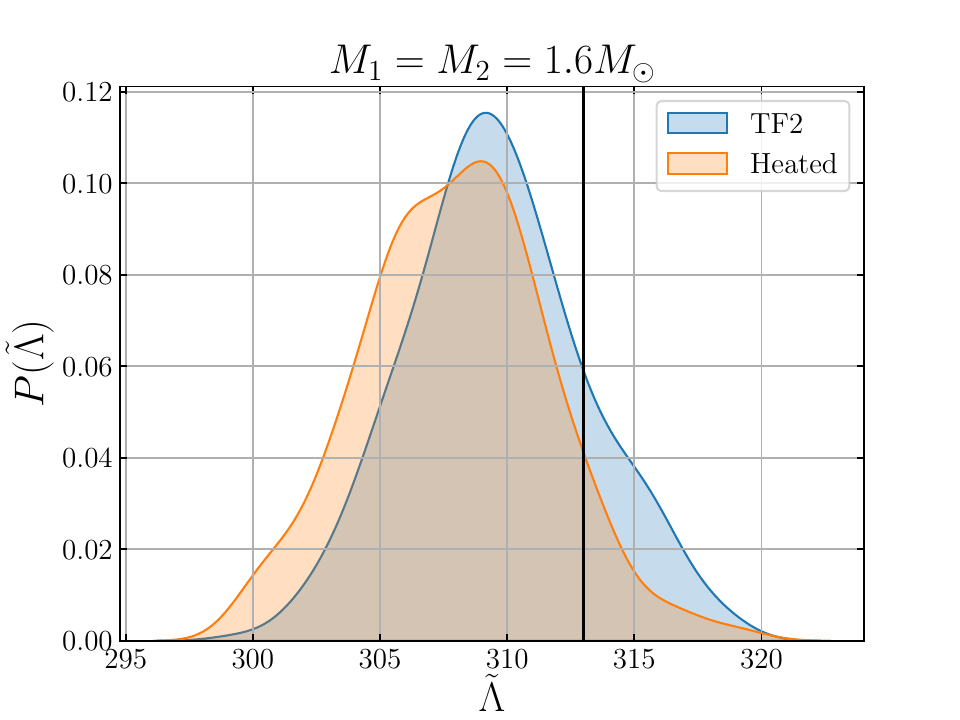}
\end{subfigure}\hspace*{\fill}
\begin{subfigure}{0.32\textwidth}
\includegraphics[width=\linewidth]{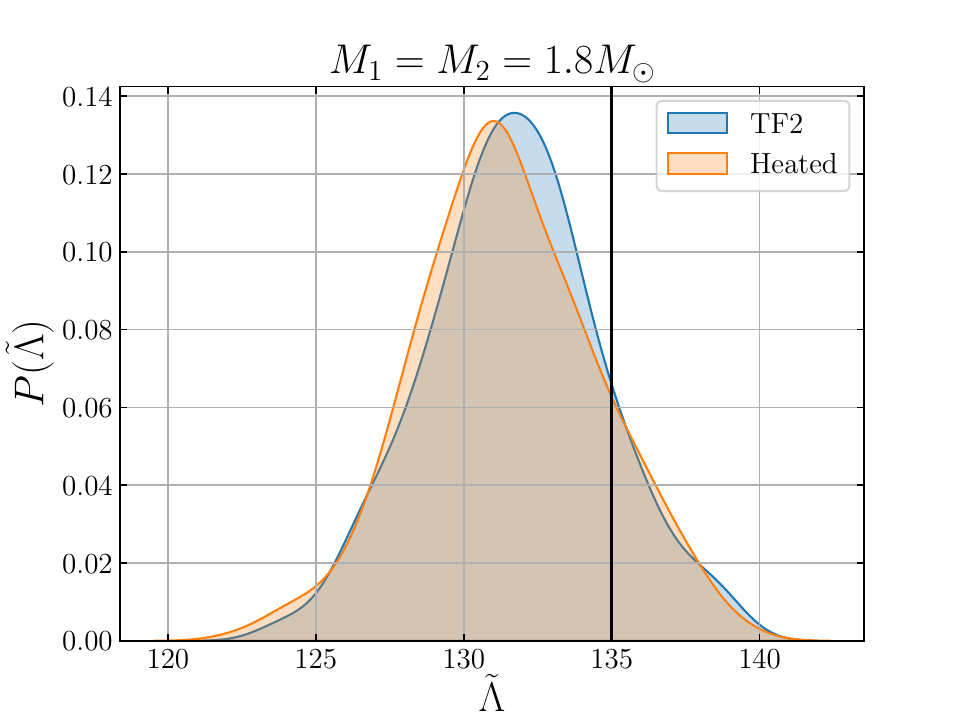}
\end{subfigure}\hspace*{\fill}
\begin{subfigure}{0.32\textwidth}
\includegraphics[width=\linewidth]{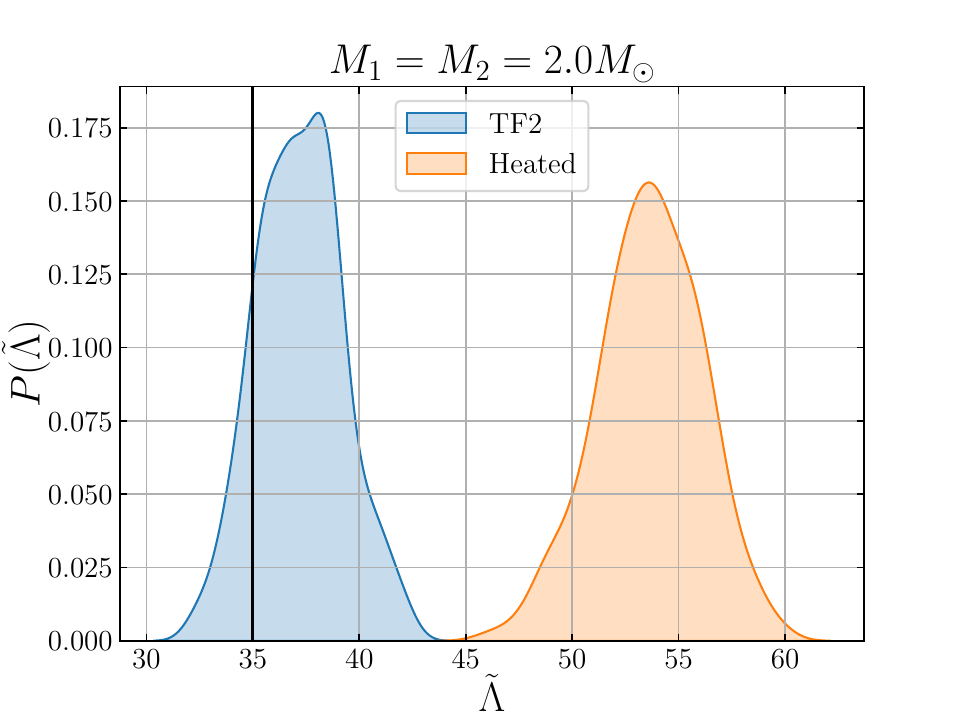}
\end{subfigure}

\medskip
\begin{subfigure}{0.32\textwidth}
\includegraphics[width=\linewidth]{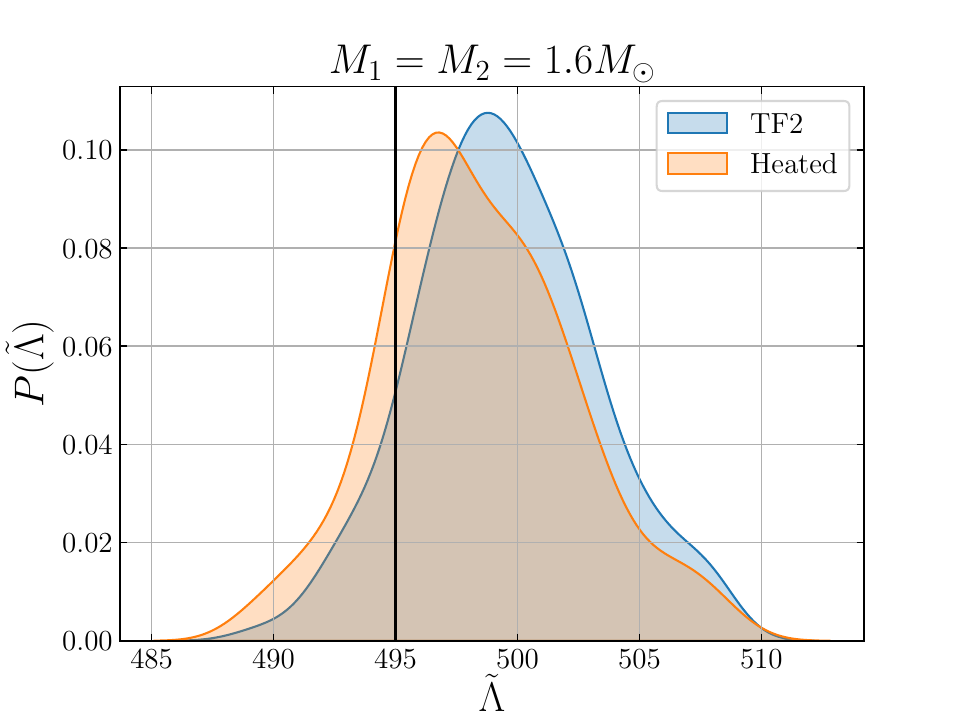}
\end{subfigure}\hspace*{\fill}
\begin{subfigure}{0.32\textwidth}
\includegraphics[width=\linewidth]{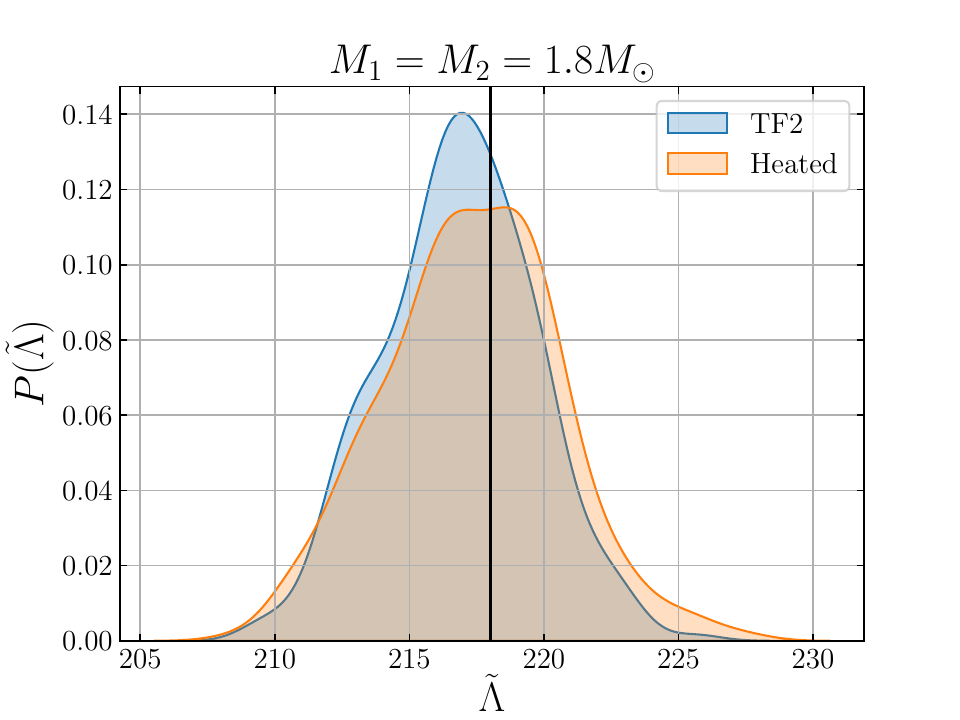}
\end{subfigure}\hspace*{\fill}
\begin{subfigure}{0.32\textwidth}
\includegraphics[width=\linewidth]{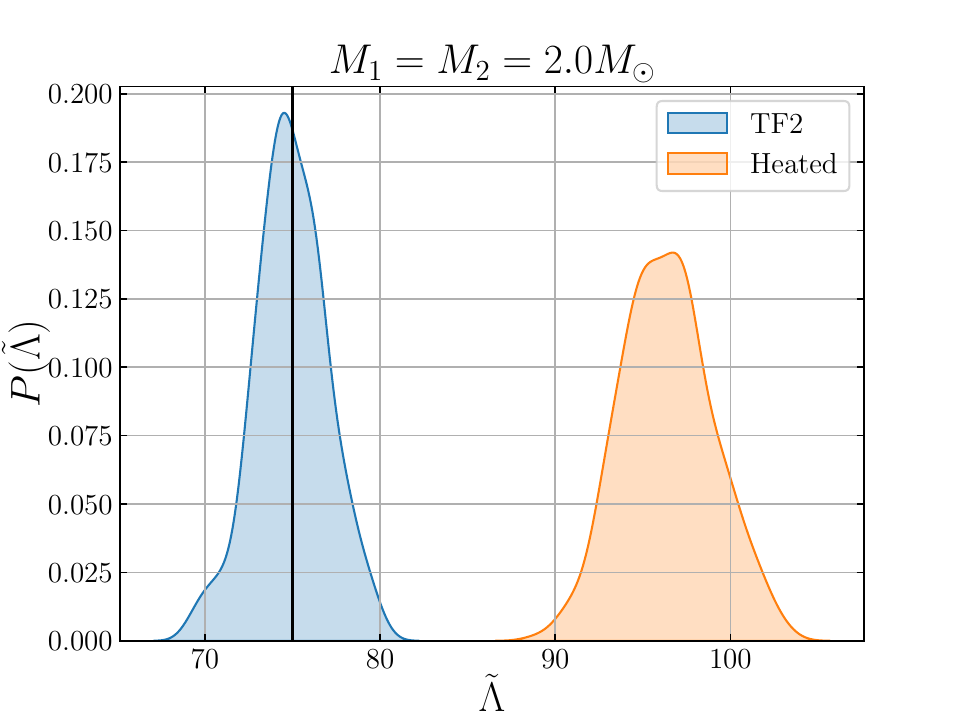}
\end{subfigure}

\medskip
\begin{subfigure}{0.32\textwidth}
\includegraphics[width=\linewidth]{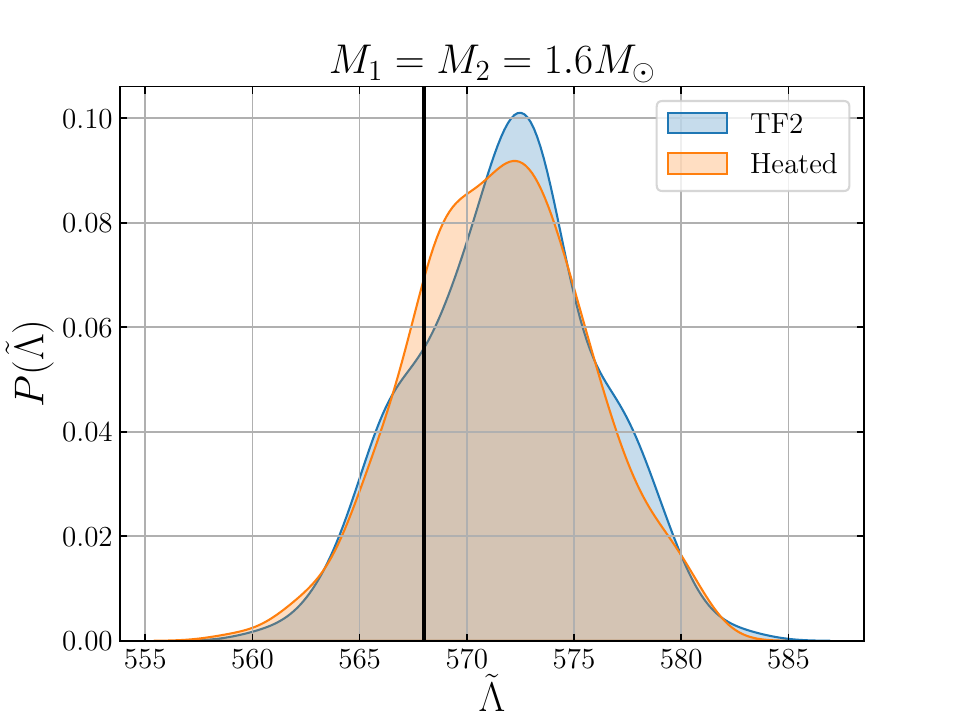}
\end{subfigure}\hspace*{\fill}
\begin{subfigure}{0.32\textwidth}
\includegraphics[width=\linewidth]{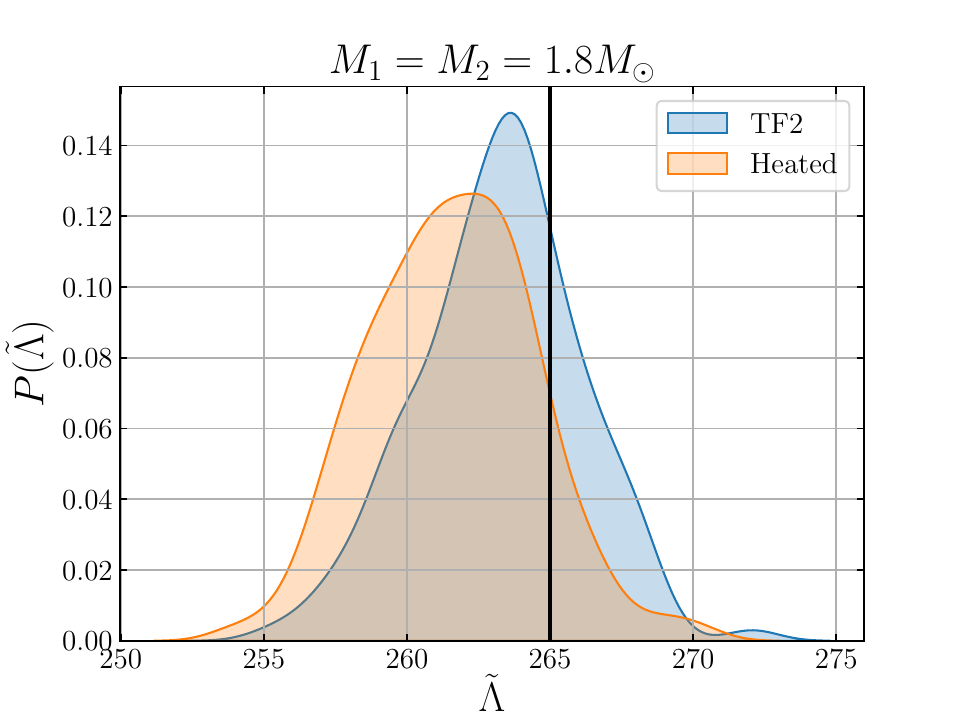}
\end{subfigure}\hspace*{\fill}
\begin{subfigure}{0.32\textwidth}
\includegraphics[width=\linewidth]{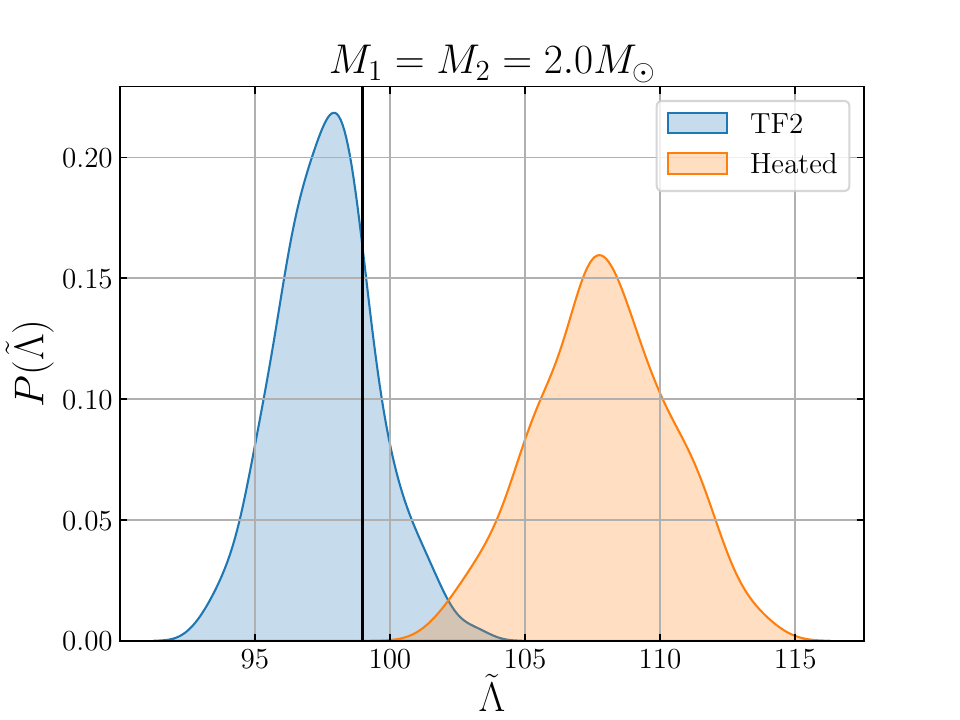}
\end{subfigure}

\caption{Injection and recovery of effective tidal deformability from single BNS event at 150 Mpc distance with CE sensitivity of equal masses for three different EOSs: a) HTZCS(upper panel) b) FSU2( middle panel) and c)NL3 (lower panel). Blue and orange posteriors show injection without (TF2) and with effects of tidal heating (Heated) respectively. Recovery is always done with the TF2 model. Black line shows the injected value.} 

\label{fig:fig_CE}
\end{figure*}

Since none of the current BNS waveforms incorporate the effects of tidal heating, we do an `Injection-Recovery' study to see how the tidal deformability recovery with current waveform models gets biased for ignoring the effects of tidal heating. We consider the simulated inspiral only frequency domain TaylorF2 (denoted by `TF2' henceforth) waveform model with 3.5 PN point particle phase, adiabatic tidal effects up to 7.5 PN~\citep{Tidal_PN,Mandal_2025} as implemented in LALSimulation~\citep{lalsuite}. We assume that the neutron stars are non-spinning and the orbits are quasi-circular, i.e., we ignore the individual spins and orbital eccentricity parameters. We also do not incorporate any dynamical tides in these waveforms as they become relevant only at high frequencies close to the merger~\citep{Prattendynamical}. Since in this work, we are mostly interested in the tidal heating effects in the inspiral phase, where also the tidal deformability effects are dominant, we focus on the inspiral waveform only and truncate the waveform at a frequency that is the minimum among the contact frequency or the frequency of the ISCO (innermost stable circular orbit). The injected waveform starts at a minimum frequency of 20Hz. We inject the signals with (denoted by `HeatedTF2') and without the extra phase shift introduced due to the tidal heating, but we always recover the signals without the tidal heating (standard TF2 waveform). We consider the two detector configurations of the third generation (3G) Einstein Telescope (ET) with ET-D sensitivity~\citep{Hild_2011} and 40-km long Cosmic Explorer~\citep{CE} detector as implemented in the software \texttt{BILBY}~\citep{bilby_paper}. We include stationary Gaussian noise based on the power-spectral densities of these detector sensitivities in our analysis. We focus on the nearby sources with luminosity distance = 150Mpc and also fix source position during the recovery(which is based upon the assumption that the BNS events can be associated with electromagnetic counterparts). The priors are Gaussian in Chirp mass ($\mathcal{M}$) with mean at injection value and $\sigma = 0.2M_{\odot}$, uniform in symmetric mass ratio ($\eta$) in the range $(0.1,0.26)$ and uniform for $\tilde{\Lambda}$ in the range ($0-5000$). We perform parameter estimation using the nested sampler dynesty~\citep{dynesty_paper} as implemented in the parameter estimation package \texttt{BILBY}~\citep{bilby_paper} for these simulated BNS events. In Fig.~\ref{fig:fig_5} and Fig.~\ref{fig:fig_CE}, we show the recovery of the parameter $\tilde{\Lambda}$ only for the three cases of different masses and three difference choices of EOS parametrizations from ~\cite{Ghosh_2024} considering that they cover the current uncertainity range of the EOS with HZTCS and NL3 being the softest and stiffest EOS considered here respectively for the detector ET and CE respectively. We see that for $1.6M_{\odot}$, even when we include tidal heating, the estimate of tidal deformability is not biased at all and is well recovered within 90\% credible of recovered posterior. For slightly heavier masses of $1.8M_{\odot}$, we start to see the slight biases ($\sim 5$ - $10\%$) in the recovery of tidal deformability for ET, but for CE we do not see any bias in this case as well. For stiffer EOS such as NL3, we also do not see much bias because the hyperon content in the neutron star cores is relatively less, leading to less dissipation.  But for the case of $M_1 = M_2 = 2.0M_{\odot}$, we see that the recovered tidal deformability when the injected signal has effects of tidal heating, is heavily biased irrespective of the EOSs and detector sensitivity, and the recovered posterior does not contain the injected value. This leads us to conclude that for high mass neutron stars than can contain significant hyperon fraction in their core, tidal heating can be significant and if not included or modelled in the current waveform models, can introduce a huge bias in the recovery of tidal deformability parameters and thus the EOS inference in the third generation gravitational wave detector era.

\begin{figure}
    \centering
    \includegraphics[width=0.48\textwidth]{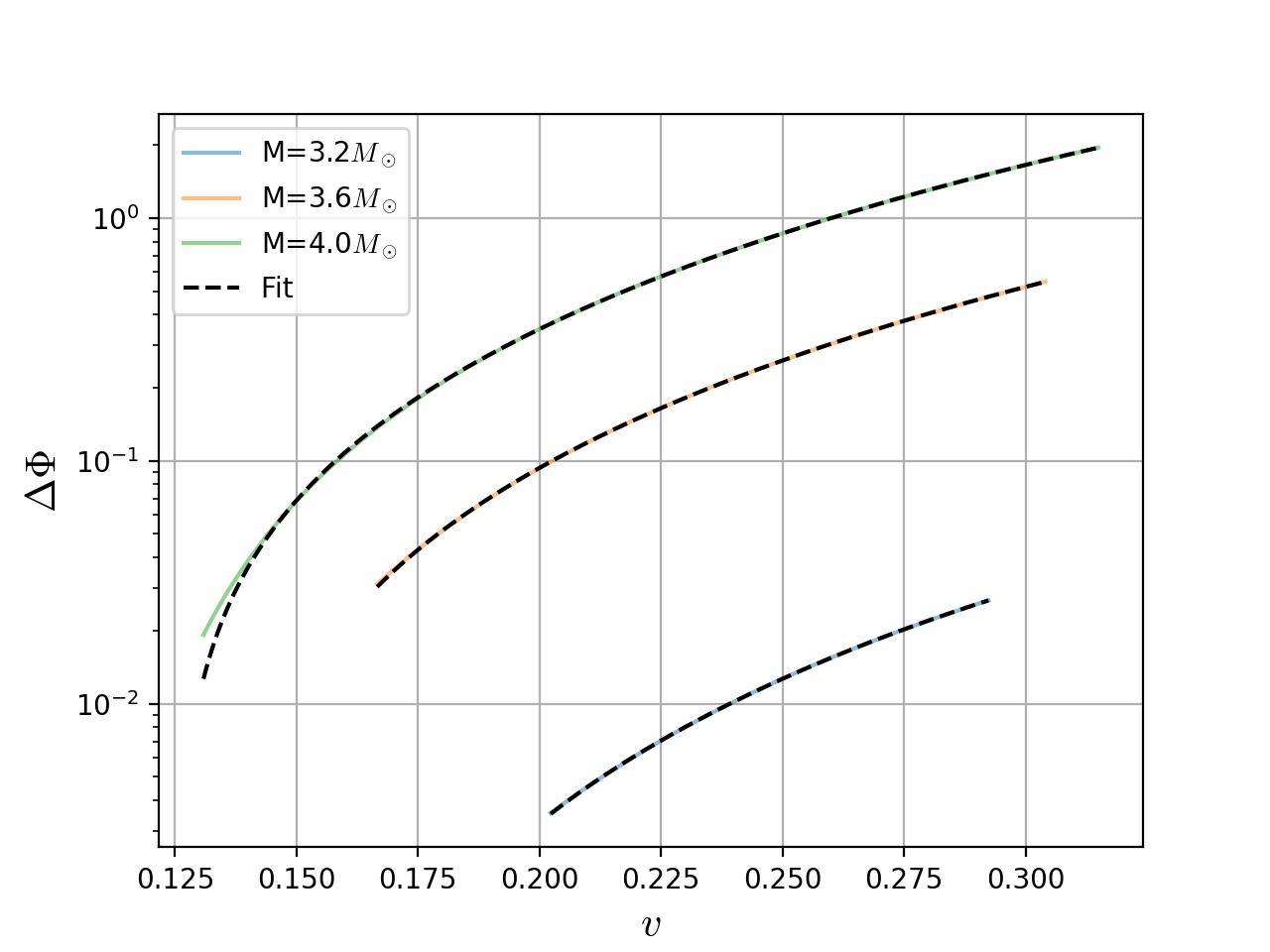}
    \captionsetup{justification=centerlast}
    \caption{\small Model of the frequency-domain dephasing due to tidal heating for three different total mass values. The solid curves show the dephasing obtained by numerically integrating Eq.~\eqref{psi}, and the individual fits with the ansatz in Eq.~\eqref{eq:ansatz} are shown by the black dashed curves.}
    \label{fig:gamma model}
\end{figure}

\section{Modeling the frequency-domain phase} \label{sec:model}

Having demonstrated that ignoring tidal dissipation can introduce systematic biases in tidal deformability estimation with current waveform models, in this section, we develop a frequency-domain model that incorporates viscous dissipation. Since we consider only non-spinning NSs, we expect the leading-order phase contribution of the viscous dissipation to appear at the 4PN order 
relative to the point-particle phase~\citep{Ripley_2023,HegadeKR:2024slr}. However, variation of the bulk viscous dissipation rate $\gamma_{\rm bulk}$ with frequency, as shown in Fig.~\ref{fig:gammavsv}, reveals that 
it has a maximum within the frequency range of interest. For a given value of the NS mass, the frequency response of $\gamma_{\rm bulk}$ differs notably before and after the maximum occurs. Although one can assume a theoretically motivated ansatz for this dissipation rate, namely, $\gamma_{\rm bulk} = Av^5/(1+Bv^{10})$, nevertheless, when considering the dependence of the bulk viscosity on the  temperature~\citep{Ghosh_2024}, it fails to capture this effect accurately over the full frequency range of interest with any phenomenological model where we expand the series around $v = 0$. One can, in principle, attempt to model the two regions around the maxima separately and connect them by imposing $C^{(1)}$ continuity, but that procedure would introduce too many parameters into the model just to incorporate one physical effect. When they are treated as free parameters in parameter estimation (PE) studies, possible degeneracies between them would hinder any meaningful conclusion. As a first step, here we implement the phase correction in frequency domain only after the maxima in $\gamma_{\rm bulk}$ occur, and set the phase correction to zero before that. This choice enables us to construct a fairly accurate model(as shown in Fig.~\ref{fig:gamma model}) with 4 new parameters, at the expense of sacrificing the phase contributions before the maxima in $\gamma_{\rm bulk}$. For high-mass neutron stars where this effect is dominant, the cut-off frequency is closer to the minimum frequency making the model accurate over most of the frequency domain. To model the numerical data for the frequency-domain phase, we consider an ansatz that contains 
the leading order 4PN and higher-order terms:
\begin{equation}\label{eq:ansatz}
    \Delta\Phi (v)=\frac{12}{128v^5}\cdot\frac{n_1 v^8\log v + n_2 v^9 + n_3 v^{10}}{1+d_1 v}\,,
\end{equation}
where 
$n_{1,2,3}$ and $d_1$ are phenomenological coefficients whose values for a $2M_\odot+2M_\odot$ BNS system considering FSU2 EOS are given in table~\ref{tab:table1}. We emphasize the fact that the parametrization used in Eq.~\eqref{eq:ansatz} is purely phenomenological in nature, and devised to analytically reproduce the numerical dephasing data. This methodology bears no foundational similarity with analytical theoretical frameworks like the post-Newtonian formalism. However, the expansion parameter $v$ remains identical to the post-Newtonian parameter, so the contributions of different coefficients thus obtained augment the baseline PN phase.
Since the leading order 4PN term is degenerate with the time of coalescence $t_c$, we consider the logarithmic term at 4PN order~\citep{Ripley_2023}. Also, the prefactor reflects the fact that we limit our investigations here to equal-mass binaries, with $\eta=1/4$.

\begin{figure}
    \centering
    \includegraphics[width=0.48\textwidth]{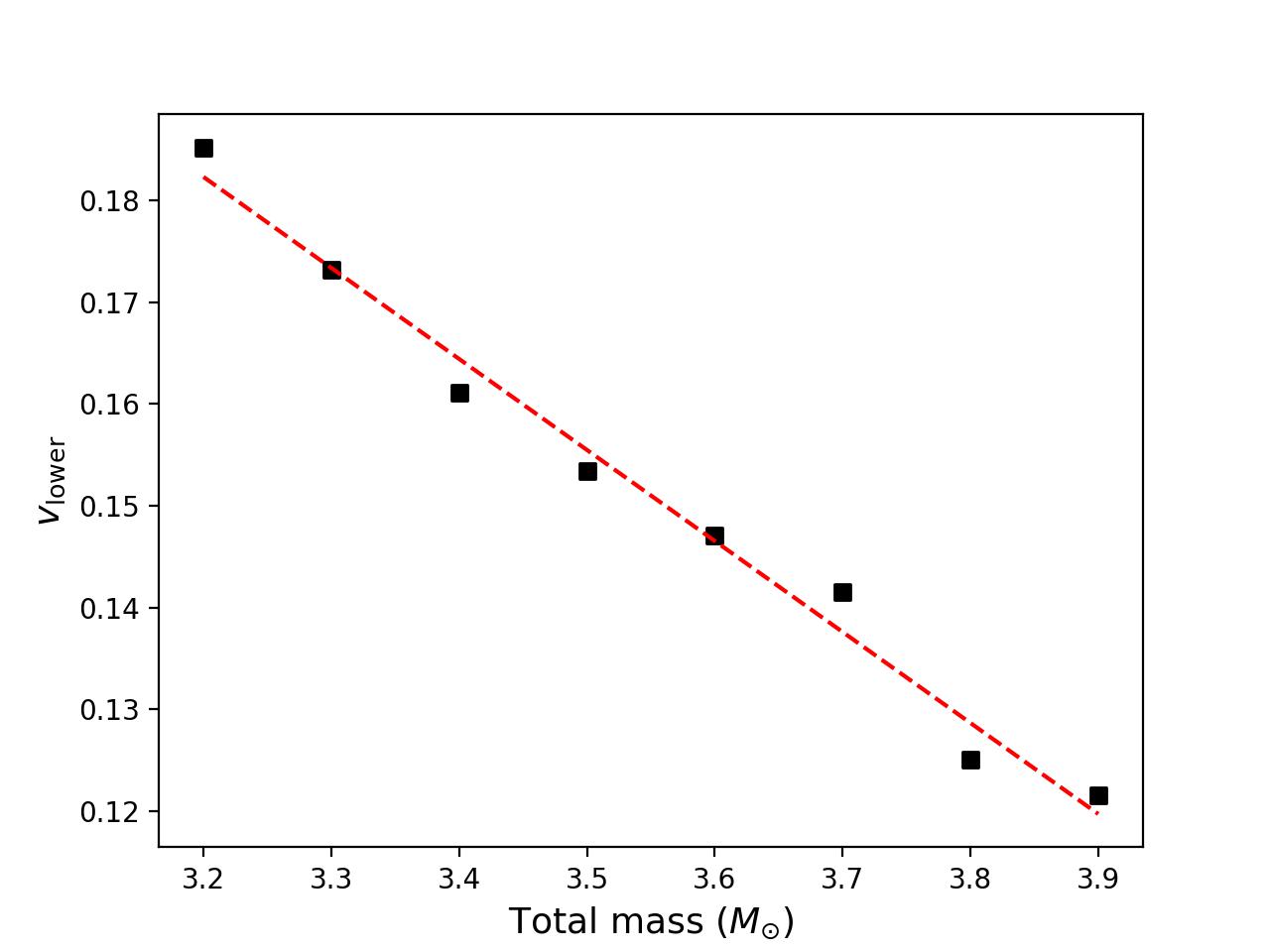}
    \caption{\small{We fit a linear ansatz with the individual maxima to model the lower cutoff. The red dashed line shows the best fit.}}
    \label{fig:lowercutoff}
\end{figure}

The lower cutoff for fitting this ansatz with data is chosen to be the frequencies at which $\gamma_{\rm bulk}$ has maxima. We fit a linear ansatz with the data for these maxima and generate a phenomenological analytical expression for the lower cutoff, given by
\begin{equation}\label{eq:cutoff}
    v_{\rm lower} = \alpha_1 + \alpha_2 (M/M_\odot)\,,
\end{equation}
with $\alpha_1 = 0.4881$ and $\alpha_2 = - 0.0893$ for the FSU2 EOS.
The fit is shown in Fig.~\ref{fig:lowercutoff}. In Fig.~\ref{fig:gammavsv} we demonstrate the bulk viscous dissipation rate as a function of $v$, and show the lower cutoff considered here. This cutoff eliminates lower frequency influence on the phase before the maxima in $\gamma_{\rm bulk}$ for all the binaries within this mass range, ensuring that the dephasing due to tidal heating appears at 4PN and higher orders. 

\begin{table*}
\begin{adjustwidth}{-3em}{}
\begin{center}
    \begin{tabular}{|c|c|c|c|c|c|c|c|c|c|}
    \hline
    Parameters & Injection & Prior distribution & Range & Unit \\
    \hline
    \multicolumn{5}{c}{} \\
    \hline
    Chirp Mass($\mathcal{M}$) & 1.76 (for $M_1 = M_2 = 2M_{\odot}$) & Gaussian & Mean = 1.76, Sigma = 0.2 & $M_{\odot}$ \\
     \hline
    Symmetric mass ratio ($\eta$) & 0.25 (for $M_1 = M_2$) & Uniform & (0.1,0.26) & Dimensionless \\
    \hline
    Effective tidal deformability($\tilde{\Lambda}$) & 75((for $M_1 = M_2 = 2M_{\odot}$)) & Uniform & (0,5000) & Dimensionless\\
    \hline
    $n_1$ & 1136 & Uniform & (10,5000) & Radian\\
    \hline
    $n_2$ & 22160 & Uniform & (10000,30000) & Radian\\
    \hline
    $n_3$ & -19951 & Uniform & (-40000,0) & Radian\\
    \hline
    $d_1$ & 14.4 & Uniform & (0,100) & Dimensionless\\
    \hline    
    \end{tabular} 
\end{center}
\end{adjustwidth}
    \caption{Choice of priors for the Bayesian posteriors presented in Fig.~\ref{fig:PE} and ~\ref{fig:Reconstruction}. }
    \label{tab:table1}
\end{table*}

\begin{figure}
    \centering
    \includegraphics[width=0.48\textwidth]{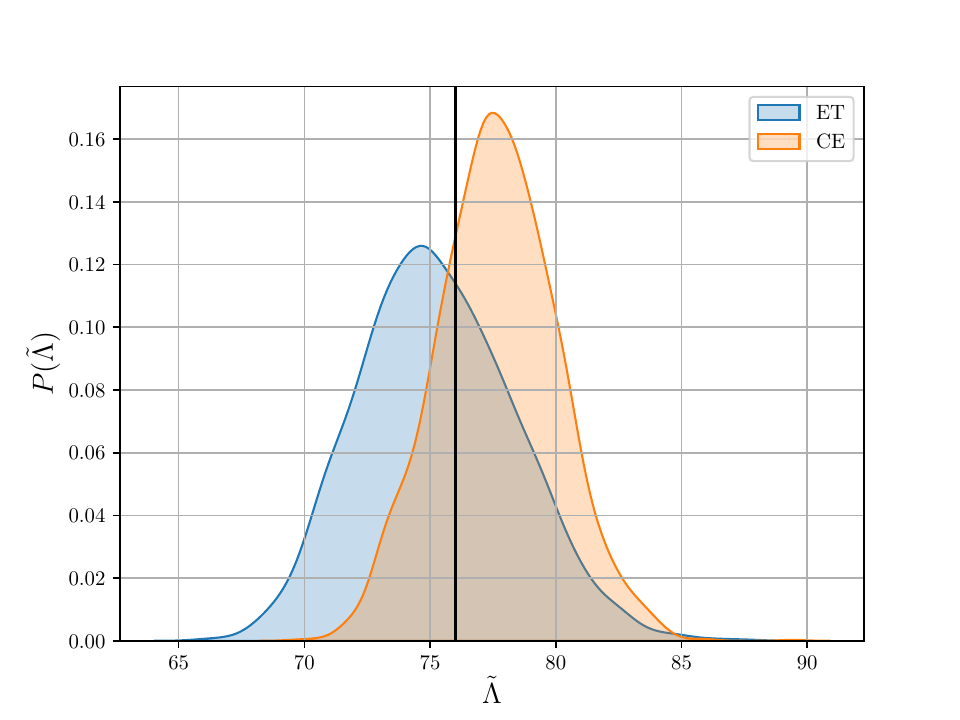}
    \caption{Recovery of tidal deformability($\tilde{\Lambda}$) for 2$M_{\odot}$ BNS system with `HeatedTF2' model}
    \label{fig:PE}
\end{figure}

\begin{figure}
    \centering
    \includegraphics[width=0.48\textwidth]{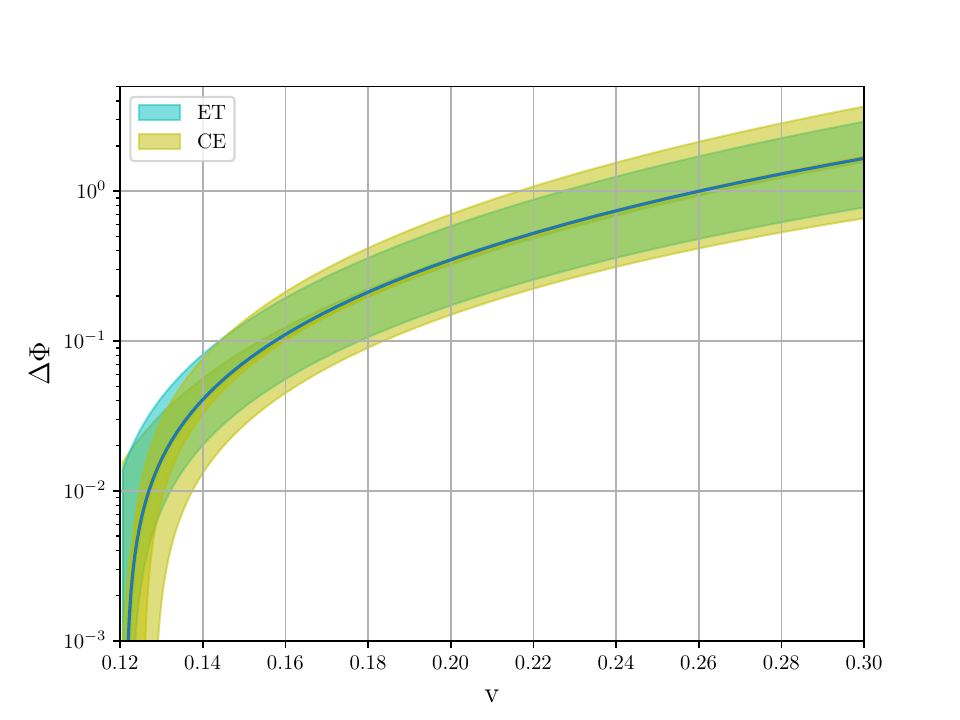}
    \caption{$1\sigma$ posterior bound of the reconstructed phase difference from the model parameters along with the injection value}
    \label{fig:Reconstruction}
\end{figure}

To test the robustness of the model and
check if we can recover the model parameters in a successful signal detection, we carry out Bayesian parameter estimation with
\texttt{BILBY}~\citep{bilby_paper}. As described earlier, we get the largest dissipation for a $2M_{\odot}$ neutron star. So, we choose an equal-mass  binary system with component masses of $2M_{\odot}$ with FSU2 EOS for the ET detector with ET-D sensitivity~\citep{Hild_2011} and 40-km Cosmic Explorer~\citep{CE}. We first inject `HeatedTF2' waveforms as described in Sec.~\ref{sec:Bias} with the tidal dissipation phase modeled as $\Phi_{\rm TH}$ in Eqn.~\eqref{eq:ansatz} with a lower frequency cutoff decided by the Eqn.~\eqref{eq:cutoff} and a higher frequency cutoff of 500 Hz. For the waveform , the starting frequency is 20 Hz and the upper cutoff frequency is taken to be the corresponding ISCO frequency. We include stationary Gaussian noise based on the power-spectral densities of the detector sensitivity in our analysis. We focus on the nearby sources with luminosity distance fixed at $150$Mpc and also fix source position during the recovery(which is based upon the assumption that the BNS events can be associated with electromagnetic counterparts). The priors for the masses, tidal deformability and the model parameters are shown in Table~\ref{tab:table1}. We do the parameter estimation using the nested sampler dynesty~\citep{dynesty_paper} as implemented in the parameter estimation package \texttt{BILBY}~\citep{bilby_paper}. Figure~\ref{fig:PE} shows the density plots of the posteriors of the tidal deformability for both the detectors. We can see that the tidal deformability is well recovered around the injection values unlike the cases when tidal heating was not modeled in the waveform(as shown in Fig.~\ref{fig:fig_5} and Fig.~\ref{fig:fig_CE}). From the recovered model parameters, we also reconstruct the phase difference that is introduced in the gravitational wave signal due to the tidal dissipation following eqn.~\ref{eq:ansatz} and plot the $1\sigma$ confidence interval of the reconstructed phase from the recovered model posterior as shown in Fig.~\ref{fig:Reconstruction}. We see that the phase difference is also well recovered around the injection value.

\section{Discussion}\label{sec:discussion}
In this paper, we have investigated the effect of viscous dissipation of tidal energy of binary neutron star systems on their gravitational waveforms. Earlier studies~\citep{Bildsten1992,Lai_1994} concluded that viscous dissipation due to the viscosity of neutron stars from nuclear matter occurs at a timescale much larger than the inspiral, and thus it does not have any observable signatures in gravitational waveforms. However,~\cite{Ghosh_2024} recently showed that if hyperons are present at the high density core of neutron stars, bulk viscosity originating from non-leptonic weak reactions involving hyperons can be much higher than the neutron star shear viscosity from $ee$ scattering and can leave detectable imprints on the GW waveforms for the next generation ground-based GW detectors. \\

In the current paper, we first briefly recapitulate the Newtonian tidal heating calculation and estimate the rate of viscous energy dissipation in the mode-sum method (the tidal perturbations are decomposed in terms of the quasi-normal modes of the star). We estimate the energy dissipated from the dominant $f$-mode oscillation due to the hyperonic bulk viscosity and estimate the additional phase contribution to the gravitational waveform using the stationary phase approximation. Comparing the dephasing thus obtained with that from the leading order static conservative tides (or tidal deformability) of neutron stars, we see that although for neutron star masses $\lesssim 1.8M_{\odot}$ the tidal deformability contributes to the phase dominantly, for heavier stars with masses $\sim 2.0M_{\odot}$ tidal dissipation contribution to the phase dominates over the tidal deformability. This behavior follows the fact that as the component neutron star masses are increased, the tidal deformability values rapidly fall down (as $\Lambda \propto M^{-6}$~\citep{Hinderer}), on the other hand, higher densities at the core of heavier neutron stars accommodate more hyperons at the core and increase the viscous dissipation. Next, we performed an injection-recovery study from a single simulated BNS event considering next generation GW detector sensitivity to show that if tidal dissipation is not modeled in BNS waveforms, it can introduce systematic biases in the recovered tidal deformability estimations, thereby biasing the equation of state inferences from GW observations that rely on accurate measurements of mass and tidal deformability. This bias is demonstrated with systems up to a maximum distance of $\sim 300$ Mpc with the third-generation GW detectors.\\

Circumventing this systematic bias entails modeling the phase correction due to the tidal dissipation and implementing it in the gravitational waveforms. Recently, efforts have been made to model this tidal dissipation or tidal lag in an effective theory of tidal responses via the parameter ``dissipative tidal deformability" that was assumed to remain constant throughout the inspiral~\citep{Ripley_2023,Ripley_2024,Saketh_2024}. However, such constant parametrization fails to capture the tidal dissipation accurately as these parameters depend on the viscosity of the neutron star matter, which is a strongly temperature-dependent quantity (refer to Fig. 3 in ~\cite{Ghosh_2024}). As a BNS system goes through the inspiral phase, the dissipated energy is converted into thermal energy, increasing the system temperature and thus changing the viscosity coefficients. For the hyperon bulk viscous dissipation, this heating was also shown to happen at a rate faster than the inspiral~\citep{Ghosh_2024}. In this Newtonian mode-sum approach, the viscous dissipation rate of the mode ($\gamma_{\rm bulk}$) characterizes the tidal dissipation. We have shown how it changes as a function of the inspiral frequency, taking into account the temperature increase due to this tidal heating. The resonance of bulk viscosity (matching of reaction rates to the perturbation timescale) is reflected in the resonance-like behavior of $\gamma_{\rm bulk}$ with the peak shifting to lower frequencies with increasing mass (and hence the dissipation). Due to this resonance, it becomes difficult to model the parameter in current state-of-the-art frequency domain BNS waveforms like `TaylorF2'~\citep{Tidal_PN} or `NRTidal'~\citep{Dietrich_2017,Dietrich_2019,Dietrich_v2,Abac_2024} based models that essentially expand the tidal phase as functions of the characteristic velocity $v$. Instead of modeling the phase correction over the whole frequency range of interest, we model it only within the frequency range $\gamma_{\rm bulk}\in [f_{\rm peak},500\text{Hz}]$ as a function of orbital velocity ($v$) starting from relative 4PN order compared to the point-particle phase with 4 additional phenomenological parameters. $f_{\rm peak}$ is the frequency corresponding to the peak of $\gamma_{\rm bulk}$. The phase is well modeled above the peak frequency with an accuracy up to 4-5\% level for all masses. Then we perform a full Bayesian parameter estimation of single simulated BNS events of $2M_{\odot}$ each for next-generation GW detectors with this model for the additional phase, and confirm that both the tidal deformability and the reconstructed phase are recovered well within their injection values. \\

This work has established that tidal dissipation in binary neutron star systems is not negligible if strangeness containing exotic matter such as hyperons are present inside their core. Tidal dissipation can thus be a smoking gun signature for their presence inside a neutron star, as there is no known source for such high viscosity from nuclear matter. Nevertheless, it is important to acknowledge certain limitations of this study, some of which are expected to be addressed in the future. 
\begin{itemize}[labelindent=0em,leftmargin=*]
    \item In our analysis, tidal dissipation in the binary neutron stars has been treated in a Newtonian mode-sum approach. Recently, the said approach was extended to include Post-Newtonian effects, but it was shown that it becomes problematic for more massive stars~\citep{Gittins_2025,yin2025postnewtonianapproachneutronstar}. The estimates presented in the current study underscore the relevance of tidal dissipation in binary neutron stars.  However, a fully relativistic theory of tidal dissipation, with accurate treatment of the underlying microphysical viscosity, would be essential for an accurate estimation of its magnitude. If such a treatment found it to be significant then the ensuing corrections to the gravitational waveform of binaries should also be incorporated.
    \item Calculation of the viscous dissipation rate ($\gamma_{bulk}$) that essentially represents the tidal dissipation has been done in a relativistic Cowling approximation. The latter is known to introduce a $20-30\%$ difference compared to fully relativistic calculations of the neutron star oscillation modes~\citep{Pradhan:2022vdf}, but the results are not expected to change qualitatively.  
    \item  Evolution of the viscous dissipation rate ($\gamma_{bulk}$) with frequency depends on the temperature evolution that was obtained as an average estimate in our earlier work~\citep{Ghosh_2024}. A detailed calculation of the temperature increase due to tidal dissipation would require a second-order perturbation theory calculation~\citep{Poisson_2011} in the relativistic perturbations of neutron stars, as well as the thermal conductivity of the various composition of neutron star from surface to the core. These should be carried out in future works for more accurate determination of the viscous dissipation rate.  
    \item The fit for the frequency domain phase due to tidal dissipation has been done with 4 phenomenological parameters. Although such a parametrization suffices to alleviate the possible biases in the estimation of $\tilde{\Lambda}$, a robust correspondence between these parameters and the physical parameters like viscous dissipation rate ($\gamma_{bulk}$) remains to be established.  
\end{itemize}

Overcoming these subtle limitations would require developing a relativistic framework for tidal dissipation incorporating fluid viscosity appropriately. In the future, consolidated efforts should be given to model the tidal dissipation in BNS waveforms more accurately based on a relativistic framework over the whole parameter space of mass and frequency, since this avenue offers a unique complimentary probe from the GW data of the out-of-equilibrium effects of dense matter at extreme conditions, whereas conservative tidal effects (static and dynamical) only probe the equation of state under equilibrium.   

\section*{Acknowledgements}
The authors thank Nils Andersson and Tim Dietrich for helpful discussions. The authors thank Bikram Pradhan for providing the calculation of $f$-mode profiles for few additional masses. The authors would like to acknowledge the usage of the IUCAA HPC computing facility, PEGASUS and  IRIDIS High Performance Computing Facility at the University of Southampton, for the numerical calculations. SG acknowledges support from STFC via grant no. ST/Y00082X/1. SB acknowledges support from the National Science Foundation under Grant PHY-2309352. We thank Jocelyn Read for reviewing the manuscript and sharing her comments.

\section*{Data Availability}

The data underlying this article will be shared on reasonable request to the corresponding author.



\bibliographystyle{mnras}
\bibliography{mnras_template} 





\bsp	
\label{lastpage}
\end{document}